\documentclass[11pt]{article}
\pdfoutput=1
\usepackage{amsmath,amssymb,epsfig,amsfonts}
\usepackage{graphicx}
\usepackage{subfig}
\usepackage[usenames, dvipsnames]{color}
\usepackage{cite}
\usepackage{multirow}
\usepackage{hyperref}
\usepackage{verbatim} 
\usepackage{enumitem}
\usepackage{rotating}
\usepackage{xcolor}
\usepackage{multirow}
\usepackage{bm}
\usepackage[normalem]{ulem}
\usepackage{cleveref}
\usepackage{slashed}
\usepackage{blkarray}
\usepackage{tikz-cd}

\usepackage{bm}

\graphicspath{ {figures/} }

\makeatletter

\newcommand{\be}{\begin{equation}}
\newcommand{\ee}{\end{equation}}
\newcommand{\ba}{\begin{aligned}}
\newcommand{\ea}{\end{aligned}}

\newcommand{\C}{\mathbb{C}}

\newcommand{\cN}{\mathcal{N}}

\newcommand{\bea}{\begin{eqnarray}}
\newcommand{\eea}{\end{eqnarray}}

\newcommand{\qline}{ \overset{ \line(1,0){30} }{\;}}

\newcommand{\R}{{\mathbb R}}
\newcommand{\Z}{{\mathbb Z}}

\newcommand{\ie}{{\it i.e.}}

\newcommand{\eg}{{\it e.g.}}
\newcommand{\Eg}{{\it E.g.}}

\def\unit{{1\kern-.65ex {\rm l}}}
\def\1{{1\kern-.65ex {\rm l}}}

\def\ds{{\displaystyle}}

\newcommand{\bX}{\bar{X}}

\def\CC{{\cal C}}

\def\CM{{\cal M}}

\def\CS{{\cal S}}
\def\CT{{\cal T}}

\def\bbF{{\mathbb{F}}}

\def\bbQ{{\mathbb{Q}}}

\newcount\hour \newcount\minute
\hour=\time \divide \hour by 60
\minute=\time
\def\now{%
\ifnum \hour<13
  \ifnum \hour=0 \advance \hour by 12 \number\hour:\else \number\hour:\fi%
     \ifnum \minute<10 0\fi%
     \number\minute%
\ A.M.%
\else \advance \hour by -12 \number\hour:%
  \ifnum \minute<10 0\fi%
  \number\minute%
  \ P.M.%
\fi%
}

\makeatother

\begin{document}

\baselineskip=18pt  
\numberwithin{equation}{section}  
\allowdisplaybreaks  






\thispagestyle{empty}

\vspace*{0.8cm} 
\begin{center}
{{\Huge Higher-Form Symmetries in 5d}}

 \vspace*{1.5cm}
David R. Morrison$^1$, Sakura Sch\"afer-Nameki$^2$, Brian Willett$^3$\\

\vspace*{1.0cm} 
{\it ${}^1$  Department of Mathematics and Department of Physics \\
University of California, Santa Barbara, CA 93106, USA}

 \vspace*{.5cm} 
{\it ${}^2$ Mathematical Institute, University of Oxford, \\
Andrew-Wiles Building,  Woodstock Road, Oxford, OX2 6GG, UK}\\

 \vspace*{.5cm} 
{\it ${}^3$ Kavli Institute for Theoretical Physics\\
University of California, Santa Barbara, CA 93106, USA}\\

\vspace*{0.8cm}
\end{center}
\vspace*{.5cm}

\noindent
We study higher-form symmetries in 5d quantum field theories, whose charged operators include extended operators such as Wilson line and 't Hooft operators. We outline criteria for the existence of higher-form symmetries both from a field theory point of view as well as from the geometric realization in M-theory on non-compact Calabi-Yau threefolds. A geometric criterion for determining the higher-form symmetry from the intersection data of the Calabi-Yau is provided, and we test it in a multitude of examples, including toric geometries. We further check that the higher-form symmetry is consistent with dualities and is invariant under flop transitions, which relate theories with the same UV-fixed point. We explore extensions to higher-form symmetries in other compactifications of M-theory, such as $G_2$-holonomy manifolds, which give rise to 4d $\mathcal{N}=1$ theories.

\newpage

\tableofcontents


\section{Introduction}

Higher-form symmetries \cite{Gaiotto:2014kfa} generalize ordinary symmetries, and have played an important role in uncovering refined  properties of quantum field theories.  For an ordinary symmetry, local operators are point-like and carry a charge, which can be measured by a surface operator surrounding the charge of dimension $d-1$ in a $d$-dimensional QFT.  A $q$-form symmetry generalizes this to charged operators of dimension $q$ and topological surface operators of dimension $d-q-1$.  Many useful properties of ordinary symmetries generalize to higher-form symmetries, including 
spontaneous symmetry breaking,  't Hooft anomalies, and selection rules.  This has led, \eg, to a deeper understanding of the phase structure of various QFTs, as well as new dualities, often for theories with little or no supersymmetry.  Much of this earlier work has focused on theories in four or fewer dimensions, often with interesting applications to condensed matter systems.

In this paper we will initiate the study of higher-form symmetries in 5d supersymmetric gauge theories.   
 Five-dimensional gauge theories are not renormalizable, however much evidence has been accumulated that points towards the existence of UV fixed points, which are interacting super-conformal field theories (SCFTs) that flow after perturbation to the gauge theory. 
Recently much progress has been made in the classification of 5d and 6d SCFTs . The case for the existence of such strongly-coupled fixed points relies on geometric constructions in M-theory and F-theory on non-compact Calabi-Yau threefolds, which model the entire parameter space of the gauge theory, including the strongly coupled limit.  For 5d $\mathcal{N}=1$ gauge theories, the Calabi-Yau geometries are resolutions of so-called canonical singularities, which characterize the SCFT \cite{Seiberg:1996bd,Morrison:1996xf}. The extended Coulomb branch, which incorporates both the vacuum expectation values of the scalars in the vector multiplet as well as masses of hypermultiplet matter fields, is geometrically realized in terms of the relative K\"ahler cone of the resolved Calabi-Yau geometry \cite{Intriligator:1997pq}. The compact 4-cycles, \ie, divisors, characterize the Cartan subalgebra of the gauge group, where their volume sets  $1/g_{YM}^2$.  
This geometric approach has advantages as it manifestly encodes properties of the UV-fixed point, such as the enhanced flavor symmetry \cite{Morrison:1996xf, Apruzzi:2019vpe, Apruzzi:2019opn, Apruzzi:2019enx, Apruzzi:2019kgb}, which in the geometry is encoded in the intersections of compact and non-compact divisors in the Calabi-Yau geometry. Furthermore, this approach does not 
rely on the existence of a weakly-coupled gauge theory description, and is applicable in cases where the UV fixed point does not have a Lagrangian description, such as in the case of the rank one $\mathbb{P}^2$ theory \cite{Seiberg:1996bd,Morrison:1996xf}. 

The main observation of this paper is that 5d gauge theories can have discrete higher-form symmetries, which we first study in the gauge theory description and provide criteria for their existence.\footnote{1-form symmetries in 5d and 6d SCFTs were discussed in \cite{Cordova:2016emh} (and upcoming work \cite{CDI}) using representation theory of the superconformal algebra. It was shown that there cannot be any continuous 1-form symmetries, but those works do not exclude discrete higher-form symmetries in 5d and 6d, which will be the focus of our attention.} Specifically, we will focus on the case of 1-form symmetries, which in the case of gauge theories can be understood as electric or center symmetries arising from the action of the center of the gauge group.  In general, these theories may also have 2-form magnetic symmetries, which arise when the gauge group is not simply connected.  However, we may always pass from 1-form to 2-form symmetries and back by gauging these higher-form symmetries, and so in this paper we will focus for concreteness on theories with 1-form symmetries only.  In the gauge theory language, this means we typically take the gauge group to be simply connected.  The field theoretic approach is complemented with an analysis in M-theory, where we find a geometric characterization of the 1-form symmetry in terms of the intersection theory on the non-compact Calabi-Yau threefold.  This is shown to agree with the 1-form symmetry computed from the gauge theory description, whenever such a formulation exists, but is applicable more generally, \eg, to the rank one $\mathbb{P}^2$ theory, which we show to have a $\mathbb{Z}_3$ 1-form symmetry.

Although the computation of the 1-form symmetry relies on the resolved geometry, we anticipate that the 1-form symmetry will be present also at the UV fixed point, where the gauge theory description breaks down, but the geometric description still applies. This is also supported by the fact that the 1-form symmetry is invariant under so-called `UV dualities', \ie, two IR descriptions that share the same UV SCFT, have the same 1-form symmetry. 
Such dualities can be inferred geometrically by determining distinct rulings of compact surfaces in the Calabi-Yau geometry, and we show that {permuting} these leaves invariant the intersection data that determines the 1-form symmetry. The 1-form symmetry is also invariant under changes in the Coulomb branch, which do not alter the weakly-coupled gauge theory description, which in turn correspond in the geometry to flop transitions between the compact surfaces.  In the case of toric Calabi-Yau threefolds this is particularly manifest, as the 1-form symmetry depends on the structure of the non-compact divisors (the external vertices in the toric diagram), and not on the specific triangulation of the toric diagram.

In the M-theory realization, the line operators that are charged under the 1-form symmetry are realized by M2-branes wrapping non-compact curves, which are infinite mass objects whose world-lines are line operators in the theory.  However, the charges of such operators can be screened by dynamical particles, which correspond to M2 branes wrapping compact cycles .  This perspective implies that the 1-form symmetry can be characterized in terms of the  cohomology of the non-compact Calabi-Yau threefold $M_6$ relative to its boundary, $H_2(M_6,\partial M_6)$.   Specifically, we find the 1-form symmetry group, $\Gamma$, is given schematically by \footnote{We refer to section \ref{sec:5dhfs} for the precise formula, which involves the long exact sequence in relative homology.  In addition, here and in most of this paper we assume that $M_6$ does not have torsion in its homology groups, which is the case in most of the examples considered in physics.  It would be interesting to extend these arguments to the torsion case.}
\be \label{GammaQuot}
\Gamma = H_2(M_6,\partial M_6;\Z)/H_2(M_6;\Z)\,. 
\ee
The magnetic dual 't Hooft operators are  similarly understood in terms of M5-branes wrapped on non-compact divisors, which are labeled by the Poincar\'e-Lefschetz-dual homology groups.  We can also understand this formula from the viewpoint of asymptotic fluxes of the M-theory three-form, $C$.  The quotient group in (\ref{GammaQuot}) can be straightforwardly computed from the intersection matrix of compact divisors and compact curves in $M_6$, which we use to determine it in many examples.  This point of view also allows generalization to other M-theory compactifications such as higher dimensional Calabi-Yau manifolds and $G_2$ and Spin$(7)$ holonomy spaces.

All 5d SCFTs that have thus far been constructed have an origin in a 6d SCFT, by circle-compactification and mass deformations (or equivalently, holonomies in the flavor symmetry). In 6d theories, one can have non-trivial 2-form symmetries, giving rise to the ``defect group'' of the theory, which is determined by the base of the elliptic fibration in F-theory models  \cite{DelZotto:2015isa, Garcia-Etxebarria:2019cnb}. 
There are also discrete 1-form symmetries in 6d and together with the 2-form symmetry they contribute to the 1-form symmetry in the dimensional reduction to 5d. We provide evidence for this by determining the higher-form symmetries in 6d for the building blocks, namely the non-Higgsable clusters (NHCs) and conformal matter theories, and find agreement with the 1-form symmetry in 5d computed by intersection theory in the resolved Calabi-Yau threefold. 

We begin the paper with a brief review of higher-form symmetries in section \ref{sec:hfs}, where we discuss also the 5d gauge theory description of 1-form symmetries. In section \ref{sec:higherformmtheory} we find a characterization of the 1-form symmetry in terms of the geometry of the M-theory realization, and find an explicit formula in terms of the intersection theory on the Calabi-Yau threefold. In section \ref{sec:5d} we apply this formalism to a large class of 5d theories, starting with the $\mathcal{N}=2$ theories, and pure Super-Yang Mills. We furthermore give a prescription for how to determine the 1-form symmetry for any collection of glued compact 
surfaces which are (blowups of) Hirzebruch surfaces. This allows us to check numerous UV dualities in section \ref{sec:Dualities}, which are shown to be consistent with the 1-form symmetry. Another large class of 5d SCFTs are obtained from toric Calabi-Yau threefolds, which we discuss in section \ref{sec:Toric}. 
To provide a more systematic framework that applies to all 5d SCFTs, it is key to determine the relation between the higher-form symmetry in 6d and in 5d, which we provide in section \ref{sec:6dto5d}. The higher-form symmetry for NHCs and conformal matter theories are computed and shown to be in agreement with the dimensional reductions to 5d. 
In section \ref{sec:gen} we consider generalizations to other M-theory compactifications. We also discuss the relation of the 1-form symmetry to the fluxes of the three-form in M-theory, and the associated surface operators.  We close in section \ref{sec:Conclusions} with conclusions and an outlook to future directions. The appendices contain computational details. 

While we were completing this paper we were informed of related work to appear in \cite{Albertini:2020mdx} after one of us presented our results in \cite{Webinar}.


\section{Higher-Form Symmetries in Gauge Theories}
\label{sec:hfs}

In this section we briefly review higher-form symmetries in general, and then focus on the case of most interest in this paper, which are 5d $\cN=1$ gauge theories.  

\subsection{Generalities about Higher-Form Symmetries}
\label{sec:hfsgen}

Higher-form symmetries \cite{Gaiotto:2014kfa} can be defined in terms of a set of topological surface operators, which obey various axioms.   We refer to  \cite{Gaiotto:2014kfa} for details, and only sketch some general features, mostly to fix notations.  

A $q$-form symmetry with group $\Gamma$ implies the existence of a set of $(d-q-1)$-dimensional topological surface operators labeled by an element of $\Gamma$, which are called the ``charge operators.''  These fuse according to the group law of $\Gamma$.  They may have non-trivial correlation functions with operators of dimension $q$ with which they are linked in spacetime, 
and these operators may be decomposed into representations of $\Gamma$.  The case of $q=0$ corresponds to ordinary symmetries, where the charge operators measure the charge of all local operators contained in the surface.  For $q>0$, $\Gamma$ must be abelian.  If the group $\Gamma$ has a continuous part then  there is a higher-form conserved current, but if $\Gamma$ is purely discrete,  there is not. We will mostly focus on the discrete case. 

Gauge theories provide natural examples of theories with higher-form symmetries.  Let $G$ be the gauge group of the theory, which we may assume has some charged matter content transforming in a (generally reducible) representation, $\bm{R}$.  Let $Z_G$ be the center of $G$, and let $\Gamma_e$ be the subgroup of $Z_G$, which acts trivially on $\bm{R}$.  If this is non-trivial, then we say the theory has an ``electric'' 1-form symmetry with group $\Gamma_e$.  Similarly, if $G$ is non-simply connected, with $\pi_1(G) \cong \Gamma_m$, then the theory has an ``magnetic'' $(d-3)$-form symmetry with group $\Gamma_m$.  The charged operators of the electric symmetry are Wilson loops, which are always 1-dimensional, and those of the magnetic symmetry are 't Hooft operators, which have dimensions $d-3$ in $d \geq 3$ spacetime dimensions.

Given a discrete $q$-form symmetry with group $\Gamma$, we may ``gauge'' this symmetry,\footnote{Here we assume the symmetry is not anomalous. We will discuss anomalies shortly. } which produces a new theory with a ``dual'' $(d-q-2)$-form symmetry, $\widehat{\Gamma}$, where $\widehat{\Gamma} \equiv \text{Hom}(\Gamma,U(1))$ is the Pontryagin dual group.   Gauging the dual group returns us to the original theory.  More generally, we may gauge a subgroup $H$ of $\Gamma$, which produces a theory with  $q$-form symmetry $\Gamma/H$ and  $(d-q-2)$-form symmetry $\widehat{H}$.  For example, given a gauge theory with simply connected group $G$ and a $1$-form electric symmetry $\Gamma_e$, gauging this symmetry gives rise to the theory with gauge group $G/\Gamma_e$, which now has a $\Gamma_m= \pi_1(G/\Gamma_e) \cong \Gamma_e$ $(d-3)$-form symmetry.

If we deform the action of a QFT by a relevant local operator, this will break ordinary (0-form) symmetries the operator is charged under.  However, there may also be accidental symmetries in the IR.  In the case of higher-form symmetries ($q>0$), the former effect does not happen, since local operators are not charged under such a symmetry, and so we typically expect that the higher-form symmetry group can only increase under RG flow.  For example, integrating out  massive charged matter in a gauge theory can increase the electric 1-form symmetry, as now there can be a larger subgroup of the center acting trivially on the remaining matter.  We will see this monotonicity property is satisfied in examples below.

\subsection{5d $\cN=1$ Gauge Theories}
\label{sec:5dhfs}

Let us now specialize to the case of gauge theories with $d=5$, and in particular to theories with $\cN=1$ supersymmetry, which will be the main focus of  this paper.  First, consider a Yang-Mills theory with a simple gauge group, $G$, which we take to be  simply connected.\footnote{As discussed above, we always have the option to take the gauge group to be simply connected, which may lead to 1-form electric symmetries.  Gauging some or all of these symmetries gives rise to the possible non-simply connected versions of the gauge group, and these will have 2-form magnetic symmetries.}   Let $Z_G$ be the center of the group.  If there are no further matter representations or interaction terms, the theory has an electric 1-form  symmetry with group $Z_G$ \cite{Gaiotto:2014kfa}.   More generally, if we include matter in some (possibly reducible) representation, $\bm{R}$, then the 1-form symmetry is broken to the subgroup, $\Gamma$, of $Z_G$ which acts trivially on this representation.  For example, for $SU(N)$ with fundamental hypermultiplets, there is no $1$-form symmetry, while for $\text{Spin}(N)$ with vector hypermultiplets, there is a $\Z_2$ symmetry.

We may characterize this in general as follows. We start with the center of the gauge group, $Z_G$, which is isomorphic to the quotient of the coweight lattice by the coroot lattice of the Lie algebra
\be
Z _G= \Lambda_{\text{cw}}/\Lambda_{\text{cr}} \,.
\ee
The center is furthermore isomorphic to the automorphism group of the extended Dynkin diagram associated to $G$\cite{IwahoriMatsumoto}.  Each irreducible representation $\bm{R}$ of $G$ determines a character, $\chi_{\bm{R}}$ of $Z_G$.  The set of all such characters generate a subgroup, $H$, of $\widehat{Z}_G$, the group of characters of $Z_G$, \ie, if the matter representation decomposes into irreducible representations $\bm{R} = \oplus_i \bm{R}_i$, then
\be H = \langle \; \chi_{\bm{R}_i}  \; | \;  {\bm R}_i  \;  \text{is a matter irrep} \; \rangle \;. \ee 
Then the 1-form symmetry, $\Gamma$, preserved by the matter is the annihilator of this subgroup, \ie,
\be \Gamma =  A(H) \equiv \{ \; z \in  Z_G \; \big| \; \chi(z) = 1 , \; \forall \chi \in H \; \} \;. \ee
For the simple gauge groups we list the centers in table \ref{tab:GZG} and the transformations of some of the representations;
see, \eg, Theorem 23.12 in \cite{FultonHarris} and \cite{gal-78}.
For $E_n$ the representations refer to $\bm{F} = (1,0, \cdots, 0)$, $\bm{C}= (0, \cdots, 0, 1, 0)$, $\bm{S}= (0, \cdots, 0, 1)$, where the roots are labeled $\alpha_1 \cdots,  \alpha_{n-1}$ and $\alpha_n$ attaches to $\alpha_{n-3}$.
For $E_6$ this is $\bm{F}= \bm{27}$, $\bm{S}= \bm{78}$ and $\bm{C}= \overline{\bm{27}}$, whereas for 
$E_7$, $\bm{F}= \bm{56}$ and $\bm{S}= \bm{912}$ and $\bm{C}= \bm{133}$.

\begin{table}
$$
\begin{array}{|c||c||c|}\hline
G & Z_G & \text{Charge of representations under $Z_G$} \cr \hline\hline
A_{N-1}& \mathbb{Z}_N & q(\bm{F}) = 1 \cr \hline
D_{2N} &  \mathbb{Z}_2^2 & q(\bm{F})= (1,1)\,,\ q(\bm{S})= (1,0) \,,\ q(\bm{C})= (0,1)  \cr \hline
D_{2N+1}& \mathbb{Z}_4 & q(\bm{F})= 2 \,,\ q(\bm{S})= 1 \cr \hline
B_{N} & \mathbb{Z}_2 &   q(\bm{F})= 0 \,,\ q(\bm{S})= 1 \cr \hline 
C_N & \mathbb{Z}_2 &  q(\bm{F})= 1  \cr \hline  
E_6 & \mathbb{Z}_3 & q(\bm{F})= q(\bm{C})=1\,,\  q(\bm{S})=0 \cr \hline 
E_7 & \mathbb{Z}_2 &q(\bm{F})= q(\bm{S})=1\,,\  q(\bm{C})=0 \cr \hline
G_2, F_4, E_8 & 1& - \cr \hline 
\end{array}
$$
\caption{Lie groups $G$ and their centers $Z_G$, as well as the charges of some of the representations. We will always consider the simply-connected form of the gauge group.  \label{tab:GZG}}
\end{table}

This applies also to semi-simple gauge groups.  For example, consider a quiver
\be SU(N_1) \;\; \qline \;\; SU(N_2) \;, \ee
with bifundamental matter connecting the two gauge groups.  Without this matter, the 1-form symmetry would be the center, $\Z_{N_1} \times \Z_{N_2}$.  The matter is charged under the center  as
\footnote{Here we denote characters of $\Z_N$ by the elements of $U(1) \cong \R/\Z$ to which they map the generator of $\Z_N$.} 
$(\frac{1}{N_1},-\frac{1}{N_2})$,
which generates a $\Z_{N_1 N_2/d}$ subgroup of  $\widehat{Z}_G$, where $d=(N_1,N_2)$ is the greatest common divisor, and the 1-form symmetry is given by the annihilator, $\Z_d$, generated by $(N_1/d,N_2/d)$.

\subsubsection*{Chern-Simons Terms and Theta Angles}

The above prescription determines the 1-form symmetry for all gauge groups besides $SU(N)$, for $N\geq 3$, and $Sp(N)$.  However, in these cases, there are additional interaction terms we may include that can affect the 1-form symmetry, namely, the Chern-Simons and theta angle terms, respectively.  These also interact with the matter content as, \eg, integrating out charged matter can shift the effective CS term.  Thus we must take care with how we specify these terms in the presence of charged matter.

First consider a pure $SU(N)$ gauge theory, for which we include a level $k$ Chern-Simons term.  Then we claim the $1$-form symmetry is broken from $\Z_N$ to $\Z_{\text{gcd}(N,k)}$.  This follows from the  fact that an instanton particle in the presence of such a CS term gains an effective charge of $k$ under the $\Z_N$ center of the group, and  so breaks the 1-form symmetry as would a charged matter field.  Similarly, we claim for $Sp(N)$, a non-trivial theta angle (\ie, $\theta=\pi$) breaks the $1$-form symmetry from $\Z_2$ to the trivial group.\footnote{Another way to see that the symmetry is broken is to note that if we place the $SU(N)/\Z_d$ theory on a spacetime $M_4 \times S^1$, the presence of fractional instantons means the CS term is not invariant under large transformations along the $S^1$ direction.  A similar argument holds  for the theta angle in $Sp(N)$ theories.}

Next let us include matter.  First, consider an $SU(N)$ theory with fundamental matter.  We may integrate out the matter with a large positive or negative mass, and the theory we obtain after doing so will have some CS levels differing by $1$, say, $k$ and $k+1$, respectively.  Normally this is described by saying the level of the theory with the fundamental matter is $k+\frac{1}{2}$, and integrating it out leads to a shift by $\pm \frac{1}{2}$.  However, the actual CS term appearing in the action must be integer quantized, and the effective shift by $\frac{1}{2}$ from the matter is an effect of the fermion determinant, which is not really a half-integer CS term.  This is discussed in \cite{Closset:2018bjz} (see also \cite{Closset:2018ghr} for an analogous discussion in the 3d case).  

For the purposes of determining the 1-form symmetry, it is always the ``bare,'' integer-quantized CS term appearing in the action that is relevant.  To make this manifest, we will take the ``$U(1)_{-1/2}$'' convention of \cite{Closset:2018bjz}, which means that, \eg, for $SU(N)_k$ with a fundamental hypermultiplet, integrating out the CS term with a large positive (respectively, negative) mass leads to the pure $SU(N)_k$ (respectively, $SU(N)_{k-1}$) theory.  More generally, integrating out a representation $R$ with positive mass does not shift the CS level, while doing it with negative mass may shift the CS level by an integer, $-A_R$, given by the cubic Casimir of the representation.  For example, for matter in the $m$th antisymmetric power of the fundamental representation, $R=\Lambda^m$, we have \cite{Bhardwaj:2019ngx},
\be A_R = \binom{N-2}{m-1} \frac{N-2m}{N-2}  \;. \ee

A useful way to characterize the effect of the CS term is to assign it a character in $\widehat{Z}_G$, as we did for the matter above.  Namely, to each  $SU(N)$ factor with $N>3$, and with a CS term $k$, we assign the character $\chi_k = \frac{k}{N}$.  Then the 1-form symmetry is the annihilator of the subgroup of $\widehat{Z}_G$ generated by the matter characters and all the CS characters,
\be \Gamma = A(H), \;\;\; H = \langle \chi_R, \chi_k \rangle \;. \ee

However, our choice of $U(1)_{-1/2}$ quantization is an arbitrary choice, and we must check that the 1-form symmetry is independent of this choice.  In effect, this means we must show 1-form symmetry is unchanged under shifts of the form
\be k \rightarrow k - A_R  \;, \ee
where $R$  is any representation present in the theory.  For the corresponding characters to generate the same group, $H$, it suffices to show that
\be \chi_R  \; | \; \chi_{A_R} \;. \ee
Namely, if this is true, then
\be \langle \chi_R, \chi_k \rangle = \langle\chi_R, \chi_{k-A_R} \rangle \;, \ee
and so the 1-form symmetry is the same for both conventions.  For example, for the $m$th antisymmetric representation, $\Lambda^m$ of $SU(N)$, which has $\chi_R = \frac{m}{N}$, it suffices to show that:
\be (m,N ) \; \bigg| \; \binom{N-2}{m-1} \frac{N-2m}{N-2} \;. \ee
We have checked this for many values of $N$ and $m$ and this appears to hold, as well as the analogous statement for more general representations.  It would be interesting to prove this in full generality.

Let us consider again the quiver:
\be SU(N_1)_{0} \;\; \qline \;\; SU(N_2)_{0} \;, \ee
where we now make explicit the CS  terms.  Above we saw this has $1$-form symmetry $\Z_d$, where $d=(N_1,N_2)$.  But had we taken the opposite convention for the CS levels, we would instead find:
\be SU(N_1)_{-N_2} \;\; \qline \;\; SU(N_2)_{-N_1} \;. \ee
Now the CS levels break the $1$-form symmetry in  each node to $\Z_d$ before accounting for the matter, and then  the matter breaks this to the diagonal subgroup.  Thus we indeed obtain the same answer using either convention.

Similar considerations apply for the theta angle in $Sp(N)$ theories.  Here we take the convention that integrating out fundamental matter with a positive mass does not affect the theta angle, while doing it with a negative mass shifts it by $\pi$.  

Finally, let  us comment on anomalies.  We claim that the above 1-form symmetries are non-anomalous, and so may be freely gauged to pass between the electric 1-form symmetries and magnetic 2-form symmetries, or equivalently, between simply and non-simply connected versions of the gauge group.  Namely, a  $q$-form symmetry has an anomaly when the charge operators are themselves charged under the symmetry.  For this to be possible, they must have the same dimension as the charged operators, \ie, $q=d-q-1$.  For example, this can happen for 1-form symmetries in 3d, and the anomaly is typically contributed by 3d Chern-Simons terms.  In contrast, 1-form symmetries cannot have such  an anomaly in 5d, and we claim that 5d Chern-Simons terms do not just contribute an anomaly, but actually break the 1-form symmetry.  We can understand this difference from the fact that, in 5d, there are dynamical particles (the instanton particles) that gain a charge in the presence of the 5d Chern-Simons terms, and so break the symmetry, while an analogous process does not occur in 3d.  We note, however, that there could, in principle, be an anomaly for 2-form symmetries in 5d.  However, all 2-form symmetries we encounter in this paper arise by gauging a non-anomalous 1-form symmetry and so are not anomalous.

In the next section, we will determine the 1-form symmetry of 5d $\cN=1$ theories engineered from CY3 compactifications in terms of the topology of the Calabi-Yau manifold. In many cases, these theories have effective descriptions as 5d gauge theories, in some cases having more than one such description.  
We will check below that the above field-theoretic prescription and the geometric prescription for the 1-form symmetry agree.

\section{Higher-Form Symmetries in 5d M-theory Compactifications}
\label{sec:higherformmtheory}

In this section we derive the higher-form symmetry of QFTs that arise from compactification of M-theory on non-compact manifolds.  In this section we will focus on the case of 5d $\cN=1$ theories arising by compactification on a Calabi-Yau threefolds.  However, many of the arguments we will use are more general, and we comment on extensions to other compactifications in section \ref{sec:gen} below.

\subsection{Canonical Calabi-Yau Threefold Singularities}

To characterize a 5d $\mathcal{N}=1$ gauge theory, and its UV SCFT fixed points, we consider M-theory on a canonical Calabi-Yau threefold singularity \cite{Morrison:1996xf, Intriligator:1997pq}. A crepant resolution, \ie, one which keeps the canonical class trivial, introduces a collection of complex surfaces $S_i$, $i=1, \cdots, r$, dual to $(1,1)$-forms. These are identified in M-theory  by expanding the three-form $C_3 = \sum_{i=1}^r A_i \wedge \omega^{(1,1)}_i$, {in terms of} the Cartans 
of the gauge group, where $r$ is always the rank of the gauge group. 
The extended K\"ahler cone is identified with the extended Coulomb branch (including both vevs of scalars in the vector multiplet as well as mass terms for hypermultiplets). 

Along special loci in the Coulomb branch, the theories can have non-abelian gauge theory descriptions. In the geometry this corresponds to the existence of rulings of the surfaces $S_i$. The surfaces can be collapsed to curves of singularities along the sections of the rulings.   The gauge couplings are set by $\text{vol} (S_i) = 1/g_i^2$ so that collapsing these curves further to zero volume results in the UV fixed point. Much recent progress has been made in studying 5d gauge theories and SCFTs using this geometric approach \cite{Xie:2017pfl, DelZotto:2017pti, Jefferson:2018irk, Apruzzi:2018nre, Bhardwaj:2018yhy, Bhardwaj:2018vuu, Apruzzi:2019vpe, Apruzzi:2019opn, Apruzzi:2019enx, Apruzzi:2019kgb, Bhardwaj:2019ngx, Bhardwaj:2019jtr, Bhardwaj:2019fzv, Bhardwaj:2019xeg, Closset:2018bjz, Bhardwaj:2020kim, Eckhard:2020jyr, Bhardwaj:2020gyu}.

There are various approaches to studying the geometry of the canonical Calabi-Yau threefold singularity and its resolutions. 
As  we will see shortly, what is important in determining the higher-form symmetry is the intersection theory of compact and non-compact cycles in the resolved geometry. 
The simplest class of geometries are toric Calabi-Yau threefolds, for which the intersection matrix can be computed simply from the toric data. This is the topic of section \ref{sec:Toric}. 

A complementary approach starts with collections of compact, intersecting surfaces and characterizes when these yield 5d gauge theories and SCFTs  \cite{Jefferson:2018irk, Bhardwaj:2018yhy, Bhardwaj:2018vuu, Bhardwaj:2019fzv,Bhardwaj:2019jtr,Bhardwaj:2019xeg}. This last approach is quite useful in computing the 1-form symmetry, although it does not encode manifestly the flavor symmetry, and we discuss it in section \ref{sec:IntS}.

A top down approach, which is applicable to all 5d theories that descend from 6d by $S^1$-reduction and mass deformation, is to study the resolutions of singular elliptic Calabi-Yau threefolds that underlie the F-theory realization of the 6d theory.  Resolving these geometries results in non-flat fibrations (i.e. fibrations where the dimension of the fiber can jump due to the presence of complex surface components) \cite{Apruzzi:2019vpe, Apruzzi:2019opn, Apruzzi:2019enx, Apruzzi:2019kgb,  Apruzzi:2018nre, Eckhard:2020jyr}. In this case the intersections can be computed from the explicit resolution divisors and curves. This approach is particularly useful in studying the (zero-form)  flavor symmetry of the SCFTs, using structures called the combined fiber diagrams (CFDs). We will apply this approach in various instances to cross-check against the toric and surface approaches.

\subsection{$1$-form Symmetry from Effective Coulomb Branch Description}

In order to derive the 1-form symmetry in these theories, we will first consider {each theory} in a particularly simple phase, corresponding to an effective abelian gauge theory description on the Coulomb branch.  Namely, consider the compactification of M-theory on a  smooth, non-compact (respectively, compact) Calabi-Yau 3-fold $M_6$, giving rise to an effective 5-dimensional QFT (respectively, quantum gravity theory) at low energies.  Let us denote  by $S_{(5)}$ the spacetime of this effective theory, so that the total space is $M_6 \times S_{(5)}$. Then we find the following field content on $S_{(5)}$ \cite{Intriligator:1997pq}:

\begin{itemize}
	\item  Let  $\omega_i \in H^2_{\text{cpt}}(M_6;\R)$, $i=1,...,r$, where $r = \text{rank}( H^2_{\text{cpt}}(M_6;\R))$, run over a basis of the $2$-cocycles in de Rham cohomology with compact support.  Then we may take the following ansatz for the $C_3$ field,
	\be C_3 = \sum_i A_i \wedge \omega_i \;, \ee
	where $A_i$ is an ordinary (1-form) gauge field on $S_{(5)}$.  This leads to a $U(1)^r$ gauge group in the effective theory, with Lie algebra $H^2_{\text{cpt}}(M_6;\R)$.  
	\item These gauge fields are subject to large gauge transformations, and the coweight lattice can be identified with:\footnote{Here we assume $M_6$ does not have any torsion in its second cohomology.   Most examples studied in the literature have this property, including all examples we consider in this paper, but it would be interesting to relax this assumption.}
	\be  \Lambda_{\text{cw}} = H^2_{\text{cpt}}(M_6;\Z)  \;.\ee	
	We note for later convenience that Poincar\'e-Lefschetz duality allows us to also write
	\be \label{hd2cw} \Lambda_{\text{cw}} \cong H_{4}(M_6;\Z) \;,\ee
	so that the generators of the gauge group can also be identified with (compact) $4$-cycles, \ie, divisors of the Calabi-Yau threefold.
	\item For each compact $2$-cycle, $C \in H_2(M_6;\Z)$, we may wrap an M2 brane on $C$, leading to a particle in the effective QFT.  The charge of this particle under the gauge group generator corresponding to a divisor, $S$, is given by
	\be \label{m2charges} Q_{C}^{S} \; =\;   S \cdot_{M_6} C \; \in \Z \;, \ee
	where the RHS is the intersection number in $M_6$.  The mass of the particle is determined by the symplectic volume of $C$.
\end{itemize}

An additional ingredient in the effective description are Chern-Simons interactions, which are controlled by triple intersection numbers for the compact divisors.  We will discuss these in more detail below.

Now let us derive the $1$-form symmetry of this effective theory.  We will consider two distinct, but related, arguments.  First, consider the line operators in the theory.  Recall that wrapping M2 branes on compact $2$-cycles gives dynamical, finite mass particles.  On the other hand, wrapping them on non-compact cycles gives infinite mass probe particles, whose world lines define line operators in the theory.  These correspond to elements in the relative homology group (we will henceforth take the coefficient group to be $\Z$ unless otherwise specified),
\be H_2(M_6,\partial M_6) \equiv H^{\partial}_2(M_6) \;,\ee
where we have introduced a shorthand notation for (co)homology relative to the boundary.  These line operators can be screened by dynamical particles, and so the group labeling the distinct classes of line operators modulo screening is
\be \label{gammadef} \Gamma \equiv H^{\partial}_2(M_6)/ \text{im} f_2  \;, \ee
where we quotient by the image of $H_2(M_6)$ under the map $f_2$ in the long exact sequence associated to the pair $(M_6,\partial M_6)$,
\be \label{h2les} \cdots \longrightarrow \; H_{i}(\partial M_6) \; \overset{h_{i}}{\longrightarrow}  \; H_i(M_6)  \; \overset{f_i}{\longrightarrow} \; H^{\partial }_i(M_6) \; \overset{g_i}{\longrightarrow} \; H_{i-1}(\partial M_6) \; \overset{h_{i-1}}{\longrightarrow} \; H_{i-1}(M_6) \; \rightarrow  \cdots \;. \ee
As $\Gamma$ classifies the line operators of the theory, it is natural to conjecture that $\Gamma$ is the $1$-form symmetry group.  By exactness of the sequence \eqref{h2les}, we can also identify
\be \label{gammah1dm} 
\Gamma \cong \text{im}(g_2) = \text{ker}(h_1) \subset H_1(\partial M_6)  \;. 
\ee
That it, it is given by the $1$-cycles in the boundary, $\partial M_6$, which become trivial when included into the bulk.

For another derivation, recall that a pure $U(1)^r$ gauge theory has a $U(1)^r$ $1$-form symmetry, but this can be broken to a subgroup by charged matter.  In general, the $1$-form symmetry preserved in the presence of matter is the subgroup of the center, ${Z}_G$, of the gauge group, which acts trivially on the matter.\footnote{{Unless additional gauging is done.}}  Suppose the matter weights generate a lattice, $\Omega^*$, in $\frak{t}^*$, and let $\Omega$ be the dual lattice inside $\frak{t}$.  Then $\Omega$ contains the co-weight lattice $\Lambda_{\text{cw}}$, and the subgroup of the center acting trivially on the matter is given by
\be \label{goldefdual} \Gamma \equiv \Omega/\Lambda_{\text{cw}} \;, \ee
or, by duality,
\be \label{goldef} \Gamma \cong \Lambda_{\text{w}}/\Omega^* \;. \ee
This latter relation can also be understood as the statement that the absence of matter fields of a given weight implies there are Wilson lines that are not screened by the matter, so whose 1-form charge is non-trivial, as above.

It remains to identify the lattice $\Omega^*$ of matter weights from the topology of $M_6$.  From \eqref{hd2cw}, the weight lattice can be identified with (assuming no torsion)
\be  \Lambda_{\text{w}} = \text{Hom}(H_{4}(M_6),\Z) \cong H^{4}(M_6;\Z )  \;.
 \ee
The matter fields come from M2 branes wrapping cycles in $H_2(M_6)$, and, by Poincar\'e-Lefschetz duality, this group can also  be identified with $H^{4}_{\text{cpt}}(M_6)$.  Then, from \eqref{m2charges}, the weights in $\Lambda_{\text{w}}$ which are realized by matter are those maps which arise by taking the intersection numbers with $2$-cycles.  One can show these lie in the image of the map $\hat{f}_{4}$ fitting into the long exact sequence\footnote{Here we have identified $H^i_{\text{cpt}}(M_6)$ with $H^i_{\partial}(M_6) \equiv H^i(M_6,\partial M_6)$.},
\be \label{h4les} \cdots \;  \rightarrow  \; H^{i}_{\partial}(M_6) \;  \overset{\hat{f}_i}{\longrightarrow} \; H^{i}(M_6) \; \overset{\hat{g}_i}{\longrightarrow} \; H^{i}(\partial M_6) \; \overset{\hat{h}_{i+1}}{\longrightarrow} \; H^{i+1}_{\partial}(M_6) \; \rightarrow  \; \cdots \;. \ee
Thus we have
\be \Omega^* = \hat{f}_{4}(H^{4}_{\partial }(M_6)) \subset H^{4}(M_6) \;,\ee
and so
\be \label{gammadef2} \Gamma = \Lambda_{\text{w}}/\Omega^* = H^{4}(M_6)/\text{im}(\hat{f}_{4})\,.
 \ee
But the sequence \eqref{h4les} is precisely what is obtained by applying Poincar\'e-Lefschetz duality to the sequence \eqref{h2les}, and the quotient in \eqref{gammadef2} is isomorphic to the quotient in \eqref{gammadef}, giving another derivation of this formula.

The above derivation only strictly applies for a smooth Calabi-Yau manifolds, while the cases of most interest often have some singularities.  For example, we may associate to a singular manifold, $X$, a CFT, while its regularizations, $\bX$, may correspond to effective descriptions on its (extended) Coulomb branch.  We conjecture that the $1$-form symmetry associated to a theory, $\CT_{X}$, engineered by a singular manifold, $X$, is the same as that for any point on the extended Coulomb branch, \ie, the group $\Gamma$ determined by \eqref{gammadef} is the same for any resolution,%
\footnote{More precisely, the 1-form symmetry agrees whenever two resolutions of a singular $X$,  $\bX$ and $\bX'$, differ by geometric transitions that correspond to merely distinct Coulomb branch phases of the same gauge theory. Transitions that e.g. decouple matter hypermultiplets can change the 1-form symmetry, see section \ref{sec:Flops}.}
$\bX$, of $X$.
In subsequent sections we will present several pieces of evidence for this statement.  For example, in  section \ref{sec:Flops} we will argue the 1-form symmetry is invariant under flop transitions.  In the case of a singular geometry with a non-abelian gauge theory description, we can also compare to the field theory result of section \ref{sec:5dhfs}, and this is the  subject of section \ref{sec:5d}.  In addition, in section \ref{sec:gen} we will consider an alternative derivation of the 1-form symmetry that depends on the behavior at the asymptotic boundary of the non-compact Calabi-Yau, and so is less sensitive to the de-singularization in the interior.

\subsection{Explicit Formula from Intersections}
\label{sec:higherformmtheoryint}

To give a more explicit formula for the 1-form symmetry, let us continue to assume, for simplicity, that $M_6$ has no torsion in any of its integer homology groups.  Then we may write
\be H_i(M_6) \cong H^i(M_6) \cong \Z^{b_i} \,,
\ee
where $b_i$ is the $i$th Betti number.  For compact $M_6$, we would have $b_i = b_{6-i}$, but more generally, we have from Poincar\'e-Lefschetz duality,
\be H^\partial_i(M_6) \cong H_{\partial}^i(M_6) \cong \Z^{b_{6-i}} \;. \ee
Now the intersection pairing between $H_i$ and $H_{6-i}$ can be represented by a $b_i \times b_{6-i}$ integer matrix, $\CM_i$, namely,
\be  \label{Mdef} 
\ba
H_i(M_6) \times H_{6-i}(M_6)  & \quad \rightarrow\quad  \Z \cr 
 (\omega \in \Z^{b_i}, \gamma \in \Z^{b_{6-i}})  &\quad  \mapsto \quad  \omega^T \CM_i \gamma 
 \ea
 \,. 
 \ee
Note that $\CM_i^T = \CM_{6-i}$.  Then the map $f_i$ in the exact sequence \eqref{h2les} above is explicitly
\be f_i: H_i(M_6) \rightarrow H_i^\partial(M_6)\,,\qquad 
f(\omega \in \Z^{b_i}) = \CM_{6-i} \omega \in \Z^{b_{6-i}} \,.\ee
Thus we have
\be \Gamma = \Z^{b_{4}}/ \CM_{4} \Z^{b_2} \,. \ee
Here we recall $b_{4}=r$ is the rank of the effective gauge group.  Moreover, we have that $b_2 = r+f$ with $f \geq 0$ the rank of the flavor group.\footnote{As discussed in, \eg, \cite{Closset:2018bjz}, we may identify $f= \text{rk}(H_4(M_6)) - \text{rk}(H_2(M_6)) $ with the rank the flavor symmetry.  }  Thus $\CM_{4}$ is an $r \times (r+f)$ matrix.

To compute the quotient, it is convenient to write $\CM_{4}$ in Smith normal form, as
\be \CM_{4} = S D T \;, \ee
where $S$ and $T$ are invertible $r \times r$ and $(r+f) \times (r+f)$ integer matrices, respectively, and $D$ is diagonal,
\be \label{dmat}
D = \begin{pmatrix} 
	\alpha_1 & 0 & \cdots&  0 & 0 & \cdots \\
	0 & \alpha_2 & \cdots & 0 & 0 & \cdots \\
	\vdots & \vdots & & \vdots & \vdots &  & \\
	0 & 0 & \cdots & \alpha_r & 0 & \cdots \end{pmatrix} \;,\ee
where $\alpha_i \in \Z_{\geq 0}$ and $\alpha_i$ divides $\alpha_{i+1}$.  
Then we find,
\be \label{gammasnf} \Gamma = \bigoplus_i \Z/\alpha_i  \Z\;. \ee

Incidentally, we note the transpose of this matrix, $\CM_2=\CM_{4}^T$,  appears in the quotient,
\be \label{ainq} H^\partial_{4}(M_6)/\text{im} f_{4} \;. \ee
Namely, by a similar argument, this group is given by
\be \Z^{r+f}/ \CM_2 \Z^{r} \;.\ee
Then using the same Smith normal form decomposition above, we see this group is given by
\be \label{h4gammadef}  H^\partial_{4}(M_6)/\text{im} f_{4} \cong \Gamma \oplus \Z^f  \;.\ee
In  section \ref{sec:gen} we will understand this identification from the point of view of the asymptotic fluxes of the $C_3$ field.

\subsection{Compact $M_6$ and the Swampland}

Let us consider this formula when $M_6$ is compact.  Then $\partial M_6$ is empty, so $f_i$ is an isomorphism, and so the RHS of \eqref{gammadef} is trivial, meaning there is no $1$-form symmetry.  Equivalently, by \eqref{goldef} we see that this implies $\Omega^*=\Lambda^*$, \ie, particles of all possible charges exist in the theory.  The line operators, which are realized by M2 brane states wrapping non-compact 2-cycles in the non-compact geometry, were of infinite mass. In the compact Calabi-Yau case, such states would all be of finite mass, and break the 1-form symmetry.  Both statements are consistent with the fact that we expect, given the compactness of $M_6$, that this theory describes a quantum gravity theory, and then the 
expectation from the weak gravity conjecture (see, \eg, \cite{Banks:2010zn, Harlow:2018tng}\footnote{Recent considerations on other swampland related conjectures in 5d $\mathcal{N}=1$ have appeared in \cite{Katz:2020ewz}. {A discsussion of the fate of 1-form symmetries in theories of supergravity were discussed in \cite{Hellerman:2010fv}.}}) is that such a theory (i) has particles with all possible gauge charges, and (ii) has no global symmetry, including higher-form symmetry.  On the other hand, if $M_6$ is non-compact, we may have a non-trivial $1$-form symmetry.  This is consistent with the fact that for non-compact $M_6$, the theory is expected to be a QFT, and so may have non-trivial global symmetries.

It is tempting to conjecture a converse to the above statement, namely, that any QFT not coupled to gravity has a global symmetry.
In particular, we observe that any theory with no 0-form global symmetry has a non-trivial 1-form symmetry. 
 In rank 1 this the $\mathbb{P}^2$ Seiberg theory, which has $\mathbb{Z}_3$ 1-form symmetry; in rank 2, there are two theories with no 0-form global symmetry, which are also non-Lagrangian (with geometries $\mathbb{F}_3 \stackrel{\ell}{\cup} \mathbb{P}^2$ and $\mathbb{F}_6 \stackrel{2\ell}{\cup} \mathbb{P}^2$ \cite{Jefferson:2018irk}, or models 67 and 68 in \cite{Apruzzi:2019opn}), but have 1-form symmetries $\mathbb{Z}_5$ and $\mathbb{Z}_4$ as can be computed from the geometry. 
Finally, in section \ref{sec:Toric}, we will see there are theories that are descendants of $T_N$, which have $G_F=1$ but also non-trivial 1-form symmetry. It would be interesting to explore this observation further. 

\section{5d Gauge Theories and SCFTs from Calabi-Yau Threefolds}
\label{sec:5d}

We saw in section \ref{sec:higherformmtheory} that the higher-form symmetry of an M-theory compactification is determined by the intersection pairing of the compactification manifold.  In this section, we will consider some examples of Calabi-Yau geometries engineering 5d $\cN=1$ gauge theories, and compute the 1-form symmetry by studying their topology.  We can then compare this result to the 1-form symmetry of  the corresponding field theories, as described in section \ref{sec:5dhfs}, which provides a strong consistency check of our results.

\subsection{5d $\cN=2$ Theories}
\label{sec:n2ex}

Let us start by considering M-theory on a Calabi-Yau threefold of the form
\be X = \C^2/\Gamma_{ADE} \times T^2 \,,
\ee
where $\Gamma_{ADE}$ is a finite subgroup of $SU(2)$, with ADE classification,
and $T^2$ is given a complex structure.  This engineers the 5d $\cN=2$  SYM theory for the corresponding ADE Lie algebra.  The higher-form symmetry of this theory depends on the choice of global form of the gauge group: for the simply connected choice there is a $1$-form symmetry given by the center of the group, which is the abelianization, $\text{Ab}[\Gamma_{ADE}]$.  This theory is equivalent to the 6d $\cN=(2,0)$ theory compactified on a circle, where the radius of the circle is inversely proportional to the volume of $T^2$.  We will keep the $T^2$ at finite volume, so that we are considering a 5d rather than a 6d theory. 

Let us verify this is the 1-form symmetry predicted by \eqref{gammadef}.  Let us consider the $A_1$ case for simplicity, the others are analogous.  We first resolve the singularity, which amounts to going onto the Coulomb branch of the gauge theory, and consider
\be 
\bX = (\mathcal{O}(-2) \rightarrow \mathbb{P}^1) \times T^2 \,.
\ee
Then we find
\be H_2(\bX) = \Z^2 \,,\qquad H_2^{\partial }(\bX) = \Z \,,
\ee
where the former is generated by the $\mathbb{P}^1$ and $T^2$ cycles, and the latter by the non-compact fiber, $F$, of the line bundle.  Now, two copies of the fiber are homologous to the base, $\mathbb{P}^1$, and so the map $f_2$ in \eqref{h2les} sends $T^2$ to zero and $\mathbb{P}^1$ to twice the generator of $H_2^{\partial}(\bX) $.  Thus we find
\be \Gamma = H_2^{\partial }(\bX) /\text{im} \;f \cong \Z_2 \,,
\ee
which is the expected result.

More generally, this can also be seen from the identification of $\Gamma$ with the subgroup of $H^1(\partial \bX,\Z)$ projecting to zero in bulk.  In the present case, the boundary is $S^3/\Gamma_{ADE} \times T^2$.  The $1$-cycles on $T^2$ are non-trivial in the bulk, while the torsion cycles are trivial, and so we find
\be \Gamma = H_1(S^3/\Gamma_{ADE}) \; \cong \; \text{Ab}[\Gamma_{ADE} ]\,,
\ee
which is the expected result for this ADE gauge group in 5d.

In this case, we also compute
\be  H_4(\bX) \cong \Z \,,\qquad H_4^{\partial}(\bX) \cong \Z^2\,,
\ee
with the former generated by $T^2 \times \mathbb{P}^1$, and the latter by $T^2 \times F$ and $S^2 \times F$, which is the base of this (trivial) elliptic fibration.  The map between them sends $T^2 \times \mathbb{P}^1$ to twice $T^2 \times F$, and so the quotient in \eqref{h4gammadef} is
\be   H_4^{\partial \bX}(\bX)/\text{im}(f_4) \; \cong \; \Z \oplus \Z_2 \,.
\ee
Here the $\Z$ factor, which comes from the base, $B=S^2 \times F$, of the fibration, implies the flavor rank is $f=1$, corresponding to the $U(1)$ instanton symmetry, which corresponds in the 6d uplift to the $U(1)_{KK}$ symmetry.

\subsection{Intersecting Surfaces}
\label{sec:IntS}

A useful method for describing the local geometry of a non-compact Calabi-Yau manifold is to specify a set of intersecting complex surfaces inside the Calabi-Yau, and study their local neighborhood.  In more detail,  one considers a set of blown up Hirzebruch surfaces, $\mathbb{F}_n^b$, of degree $n$ and with $b$ blow-ups,  intersecting along curves, and consistency conditions can be derived for these to embed into a Calabi-Yau threefold. In particular the interest here is in collections of surfaces that can be collapsed to zero volume, thus describing a 5d SCFT. 

First we recall some basic properties of the blown up Hirzebruch surface\footnote{If $b=0$, we suppress the $b$ in the notation.} $\mathbb{F}_n^b$.  This has curves generated by the base, $e$, fiber, $f$, and curves $x_i$ from blown-up points,  $i=1,...,b$, which satisfy
\be  \label{hcurvint} e^2=-n, \;\;\; f^2 = 0, \;\;\; e\cdot f=1, \;\;\; e \cdot x_i = f \cdot x_i = 0, \;\;\; x_i \cdot x_j = - \delta_{ij} \;. \ee
The canonical divisor can be written as
\be \label{hK} K = -2 e - (n+2) f  + \sum_i x_i \;. \ee
Given a collection of such surfaces, $S_i$, we specify which curves they intersect along by saying $S_i$ is glued to $S_j$ by identifying the curve $C_{ij}$ in $S_i$ and $C_{ji}$ in $S_j$.  Consistency dictates that the genus is 
\be g(C_{ij}) = \frac{1}{2} (C_{ij}^2 + K_i \cdot C_{ij}) + 1= g(C_{ji}) \ee
and the Calabi-Yau condition implies that the degrees of the normal bundles have to satisfy
\be C_{ij}^2 + C_{ji}^2 = 2 g(C_{ij}) -2 \,,\ee
where $C_{ij}^2$ is the self-intersection of the curve inside $S_i$.

In many cases we can read off an effective 5d gauge theory description from the collection of intersecting surfaces.  We first recall that the generators of the gauge group, $G$, correspond to the compact divisors, $S_i$, $i=1,...,r_G$.  The fibers, $f_i$ give rise to the W-bosons when we wrap M2 branes on them, and we may define:
\be N_{ij} =- S_i \cdot f_j \,,
\ee
which is (minus) the Cartan matrix of the gauge group \cite{Intriligator:1997pq,Aspinwall:2000kf}.  Note that to compute the intersection of curves with surfaces, we use:
\be \label{sc} S_i \cdot C = \begin{cases} K_i \cdot C \,,& \quad C \; \in \; S_i \\ C_{ji} \cdot C\,, &\quad C \in S_{j \neq i} \,. \end{cases} \ee

\paragraph{Example $SU(3)_0$.}
To take a simple example of a geometry, consider the following diagram
\be \mathbb{F}_1 \;_e\qline_e \; \mathbb{F}_1\,, \ee
where the notation means that we glue the two surfaces along the base curves, $e$, in the respective surfaces.  Then the matrix $N_{ij}$ above is computed to be the Cartan matrix of $SU(3)$, and in fact this geometry generates the $SU(3)_0$ pure $\cN=1$ gauge theory.

Let us consider the 1-form symmetry of this theory using the formulas of section \ref{sec:higherformmtheory}.  A basis of the curves is given by
\be \{ e_1 ,f_1,f_2 \} \,,\ee
where we recall $e_1=e_2$ due to the gluing.  A basis of the compact divisors are given by the two Hirzebruch surfaces.  Then we compute the matrix $\CM_4$ in \eqref{Mdef} as
\be (\CM_4)_{ij} = S_i \cdot C_j = \begin{pmatrix} -1 & -2 & \hphantom{-}1 \\ -1 & \hphantom{-}1 & -2 \end{pmatrix}\,. \ee
One finds the Smith normal form of this matrix to be
\be \begin{pmatrix} 1 & 0 & 0 \\ 0 & 3 & 0 \end{pmatrix}\,, \ee
which, comparing to \eqref{gammasnf} and \eqref{h4gammadef}, implies the 1-form symmetry is $\Z_3$, and the flavor rank is $1$, which is as expected.

\paragraph{General Formula for the $1$-Form Symmetry.}

More generally, we proceed as follows.   Our goal is to compute the quotient
\be H_4^\partial(X)/\text{im} f_4 \,,\ee
where $X$ is the Calabi-Yau containing the intersecting surfaces.  We recall, by Poincar\'e-Lefschetz duality
\be H_4^\partial(X) \cong H^2(X) \cong \text{Hom}(H_2(X),\Z) \,.
\ee
To compute this group, we note that $H_2(X)$ is generated by the curves $\{e_a,f_a,x_{i,a} \}$, where $a$ indexes the blown-up Hirzebruch surfaces, $S_a = \bbF_{n_a}^{b_a}$.  There are a total of
\be N_c = \sum_a (b_a + 2) \,, \ee
such generators.  However, these are subject to the relations implied by the gluing curves, $C_{ij} = C_{ji}$.  Suppose there are $N_g$ such relations, then we may organize them into an $N_g \times N_c$ matrix, $G$.  Then we have:
\be\ba \label{h2xb}  
\ds H_2(X)  &\cong \Z^{N_c}/\text{im}\; G^T  \cr 
 \ds H^2(X) &\cong \text{ker} \; G \subset \Z^{N_c} \,.
 \ea
 \ee
In both cases these are isomorphic to $\Z^{N_c-N_g}$, but the bases above will be important in the computation.  

The second line in \eqref{h2xb} represents the space of all homomorphisms from $H_2(X)$ into the integers.  The divisors provide a special class of such homomorphisms via taking the intersection pairing.   If there are $N_d$ divisors, this gives rise to another matrix, $\Omega$ of size $N_c \times N_d$,
\be\label{OmegaDef}
 \Omega_{\alpha a} = C_\alpha \cdot S_a \;, \ee
where $S_a$ runs over the $N_d$ (blown up) Hirzebruch surfaces, and $C_\alpha$ runs over the $N_c$ generators, $e_a,f_a,x_{i,a}$, and the intersection numbers can be computed in the above basis using \eqref{sc} along with \eqref{hcurvint} and \eqref{hK}.  Then it can be shown using the gluing consistency conditions above that the image of this matrix lies inside $\text{ker}\;G$, and we have,
\be \Gamma \oplus \Z^f \;\cong \; H_4^\partial(X)/\text{im} f_4 \; \cong \; H^2(X)/\text{im}\hat{f}_2 \; \cong \; \text{ker} \; G/ \text{im}\; \Omega  \;. \ee
We may equivalently summarize this by saying we have a complex,
\be \Z^{N_d} \;  \overset{\Omega}{\longrightarrow} \; \Z^{N_c} \;  \overset{G}{\longrightarrow}  \; \Z^{N_g} \;, \ee
\ie, $G  \cdot \Omega= 0$, and $\Gamma \oplus \Z^f$ is given by the homology of this complex at the central term.  One can compute this homology purely in terms of the ``elementary divisors'' of $\Omega$ -- \ie, the diagonal entries, $\omega_j$, in its Smith normal form -- and the ranks of $G$ and $\Omega$, as
\be \label{gzfgo}  \Gamma \oplus \Z^f \; \cong \; \bigoplus_{i=1}^{r_\Omega} \; \Z/\omega_i \Z \; \oplus \;\Z^{N_c - r_\Omega - r_G} \;. \ee

In appendix \ref{app:1formsurface} we argue that the 1-form symmetry determined by the formula  above agrees precisely with  that derived in section \ref{sec:5dhfs} field theoretically.

\subsection{Invariance under Flops}
\label{sec:Flops}

The 1-form symmetry should be invariant under geometric transformations, which lead to equivalent descriptions of the 5d theory. 
One such class of transformation are the flop transitions of curves with normal bundles\footnote{In general, there are other flops as well.  There is an invariant
called the ``length'' of a flop \cite[pp.~95-96]{[CKM]} which is an integer
taking values
between $1$ and $6$, and even for the length $1$ cases we have in addition to
the $(-1,-1)$ flops the more general ones studied by Reid \cite{pagoda}.  The
most general case was studied by Koll\'ar \cite{[Kol]}, and an initial classification
was made in \cite{gorenstein-weyl}.  This classification was refined in
\cite{flopfactorization,arXiv:1709.02720}, and applied to physics in \cite{quiversMF,Collinucci:2019fnh}. } $(-1,-1)$ that are contained in the compact surfaces. Such flops correspond to changes in the gauge theory Coulomb branch phase and should not affect the global symmetries of the theory.  To show that the 1-form symmetry remains invariant we need to show that the Smith normal form of $\Omega$ (\ref{OmegaDef}) is invariant under flops. 

Let us briefly sketch the argument here; we refer to appendix \ref{app:Inv} for more details.  Consider the flop of a rational curve $C_a$, with $C_a^2|_{S_a} =-1$. Furthermore let $C_a\cdot S_b=1$ and for all other surfaces, $C_a \cdot S_c=0$. 
This corresponds to a row in the matrix $\Omega$
\be \label{ombef}	\Omega_{\text{before flop}} =\begin{blockarray}{ccccc}
	& S_a & S_b & S_c & \cdots  \\
	\begin{block}{c(cccc)}
		\vdots & \vdots  & & & \\
		C_a & -1 & 1 & 0  & \cdots  \\
		\vdots & \vdots  & &  &  \\
	\end{block}
\end{blockarray}  \;. \ee
Note that, for another divisor, $S_{a'}$, which contains the curve $C_a$, there is a corresponding row with a curve $C_{a'}$, which is identified with $C_a$, which is identical to the row for $C_a$.  In the Smith normal form of $\Omega$, this just adds a row of zeros, and does not affect the group in \eqref{gzfgo}.  After the flop, the row for $C_a$ is removed, and a new row for $C_b$ appears
\be	\label{omaf} \Omega_{\text{after flop}}=\begin{blockarray}{ccccc}
	& \hat{S}_a & \hat{S}_b & S_c & \cdots  \\
	\begin{block}{c(cccc)}
		\vdots & \vdots  & & & \\
		C_b & 1 & -1 & 0  & \cdots  \\
		\vdots & \vdots  & &  &  \\
	\end{block}
\end{blockarray}  \;,\ee
with all the other rows unchanged.  The divisors, $\hat{S}_{a/b}$ are the corresponding blown-up/down versions of $S_{a/b}$.
We see that the only effect is to change the sign of one row, which clearly does not affect the Smith normal form of the matrix, and so preserves the quotient group in \eqref{gzfgo}.  Although the flops do not change the 1-form symmetry, if in addition to flopping the curve, we also take its area to infinity (which in the field theory corresponds to decoupling the associated hypermultiplet), then the 1-form symmetry can change. 
For more general flops, we claim the effect is still to flip the sign of a row in this matrix, and so this will still preserve the group. 
From the toric perspective, which we will discuss in section \ref{sec:Toric}, the invariance under internal flops is straightforward, as the 1-form symmetry only depends on the external vertices, and does not depend on the internal vertices and the triangulation.

There are several other operations in the geometry which do not change the theory, which are described in appendix \ref{app:Inv}.

\subsection{Pure Super-Yang Mills}
\label{sec:SYM}

Pure 5d $\mathcal{N}=1$ SYM for $SU(N)_k$, for\footnote{For $SU(2)$ we should use the description in terms of $Sp(1)$ below.}  $N>2$, can be constructed from a collection of $N-1$ surfaces. There are various description: toric geometry, intersecting surfaces, and non-flat resolutions of elliptic Calabi-Yau threefolds, as well as from the homology of the five-manifold that is the boundary of the non-compact Calabi-Yau.  

All these have in common that there is an intersection of $N-1$ compact surfaces $S_i$, which have a ruling along which they can be collapsed to curves of an $A_{N-1}$ singularities. The prepotential can be computed from the triple intersection numbers in the geometry, in particular the Chern-Simons level is determined by intersecting the compact surfaces 
\be
c_{ijk}= S_i \cdot S_j \cdot S_k \,.
\ee
The surfaces $S_i$ (blown up) Hirzebruch surfaces and are glued to realize $SU(N)_k$ as follows 
\be \label{SUNkSurfaces}
\mathbb{F}_{N-2-k} \; \;_e\qline_h \; \; \mathbb{F}_{N-4-k} \;  \;_e\qline \cdots  \qline_h \; \mathbb{F}_{4-N-k}  \; \;_e\qline_h \; \; \mathbb{F}_{2-N-k}   \,.
\ee
We can apply the rules for computing the intersection matrix $\mathcal{M}_4$ in the last section. The details of this analysis are shown in appendix \ref{app:1formsurface} and we find that the 1-form symmetry is 
\be\label{ref:}
\Gamma = \mathbb{Z}_{\text{gcd}(N,k)} \,.
\ee
For $Sp(N)$ with  $\theta = N \pi \mod 2 \pi$ 
\be
\mathbb{F}_{2N+2} \; \;_e\qline_h \; \; \mathbb{F}_{2N} \;  \;_e\qline \cdots  \qline_h \; \mathbb{F}_{6}  \; \;_e\qline_{2h} \; \; \mathbb{F}_{1}\,,  
\ee
and $\theta= (N+1) \pi \mod 2 \pi$
\be
\mathbb{F}_{2N+2} \; \;_e\qline_h \; \; \mathbb{F}_{2N} \;  \;_e\qline \cdots  \qline_h \; \mathbb{F}_{6}  \; \;_e\qline_{2e+f} \; \; \mathbb{F}_{0}\,. 
\ee
The 1-form symmetry in these cases is
\be
Sp(N)_{0 \;\text{mod}\; 2\pi} :
\quad \Gamma = \mathbb{Z}_2 \qquad 
Sp(N)_{\pi\; \text{mod}\; 2\pi} :
\quad \Gamma = 1 \,.
\ee

These surface configurations arise from a resolution of a canonical singularity in certain Calabi-Yau threefolds, specifically, a class of geometries that are non-flat resolutions of the elliptically fibered threefolds associated to the $(D_k, D_k)$ conformal matter theory \cite{Apruzzi:2019vpe}.  The 5d IR description is an $SU(N)_0 + (2N+2) \bm{F}$. After decoupling the flavors entirely, which in the geometry corresponds to flopping rational curves out of the compact surfaces, this realizes
\be
SU(N)_{k}\,,\qquad k=N+2 , \cdots,  0\,.
\ee
The surface geometry is (equivalent to) the one in (\ref{SUNkSurfaces}). Note also that this class of theories have a UV dual description in terms of $Sp(N-1) + (2N+1) \bm{F}$, and after decoupling fundamental hypermultiplets, realize also the theories $Sp(N-1)_{\theta}$. 

The toric description of $SU(N)_k$ was discussed in \cite{Closset:2018bjz}.  We will discuss general toric Calabi-Yau threefolds in detail in section \ref{sec:Toric}, but let us briefly sketch the argument for this special case.  Consider the toric diagram with external vertices $\hat{v}_i = (w_i, 1)$, where 
\be
w_1 = (0,0) \,,\qquad 
w_2 = (0,N)\,,\qquad 
w_3 =  (1, a)\,,\qquad 
w_4 = (-1, b) \,,
\ee
To ensure convexity we require that $0<a+b <2N$. The CS level is identified with 
\be
k= a+b -N \,.
\ee
As we will show in section \ref{sec:Toric} the 1-form symmetry in the toric cases is determined simply by taking the matrix 
\be
A = \left( \hat{v}_1, \cdots, \hat{v}_4 \right) \,,
\ee
and computing its Smith normal form.  One again finds the 1-form symmetry is $\Z_{\text{gcd}(N,k)}$.

The last point of view is to consider the five-manifold that bounds the Calabi-Yau threefold. As discussed in (\ref{gammah1dm}) $\Gamma = H_1 (\partial M)$. For 
$SU(N)_k$ the boundary 5-manifold is the Sasaki-Einstein space $Y^{N, k}$, which are circle fibrations over $S^2 \times S^2$ with $c_1$ given by $N$ and $k$, respectively. The first homology can be computed, \eg, as explained in \cite{Sparks:2010sn}, to be 
\be
H_1 (Y^{N,k}, \mathbb{Z}) = \mathbb{Z}_{\text{gcd}(N,k)} \,.
\ee
From \eqref{gammah1dm}, this  implies the 1-form symmetry is a subgroup  of $\mathbb{Z}_{\text{gcd}(N,k)}$, and since the Calabi-Yau threefold is  simply connected, all cycles become trivial in the bulk, and so this subgroup is  in fact the entire group, giving yet another derivation that the 1-form symmetry of $SU(N)_k$ is $\mathbb{Z}_{\text{gcd}(N,k)}$.

\subsection{Dualities}
\label{sec:Dualities}

Distinct 5d gauge theories can share the same UV completions in 5d or 6d. Such theories are at times referred to as ``UV duals". 
A classic example is the duality between SQCD theories with $SU$ and $Sp$ gauge groups with fundamental matter. For example for rank 2 there is the duality 
\be
SU(3)_{1} + 8 \bm{F}\quad  \longleftrightarrow \quad  Sp(2) + 8 \bm{F} \,.
\ee
In the geometric realization these can be seen by constructing the associated rulings from a given resolved Calabi-Yau geometry. 
These UV dual gauge theories share parts of their extended Coulomb branch, and we therefore expect them to have the same 1-form symmetries. \Eg, the duality above will have trivial 1-form symmetry, from both gauge theory descriptions, due to the fundamental matter. 
At rank 2 there are no UV dualities that relate two theories with non-trivial one-form symmetry (for a complete list with all dual descriptions see \cite{Jefferson:2018irk, Hayashi:2018lyv, Apruzzi:2019opn}). However, a large class of 
dualities of 5d gauge theories were conjectured in \cite{Bhardwaj:2019ngx}. The consistency with respect to the 1-form symmetry provides a non-trivial test for these dualities. 
Here we summarize those that have a non-trivial 1-form symmetry and show how it is realized in terms of the geometry and the dual gauge theory descriptions (equation labels on the left-hand side are those in \cite{Bhardwaj:2019ngx}). The 1-form symmetries are determined from the gauge theory rules in section \ref{sec:5dhfs} and, complementarily, from the surface intersections, as we have verified in many cases by explicit computation.

\paragraph{Duality (1.62)} 
\be
SU(4)_{2} + 4 \Lambda^2  \quad \longleftrightarrow\quad  
2 \Lambda^2 -Sp(2)_{0}  -SU(2)_{0} \,.
\ee
The theories on both  sides of this duality have a $\mathbb{Z}_2$ 1-form symmetry.  Gauging this symmetry implies a new duality:

\be
(SU(4)_{2})/\Z_2 + 4 \Lambda^2  \quad \longleftrightarrow\quad  
2 \Lambda^2 -(Sp(2)_{0}  -SU(2)_{0})/\Z_2 \,.
\ee
Similar comments apply to the subsequent examples, and we will  only write the duality with simply-connected gauge groups.

\paragraph{Duality (1.88)} 
\be
Sp(2l+1)_{0}   \quad \longleftrightarrow\quad  
SU(2l+2)_{-2l-4} \,.
\ee
This duality has a $\mathbb{Z}_2$ 1-form symmetry.

\paragraph{Duality (3.74)}
The geometry of surfaces is 
\be
\begin{tikzpicture}
\node (F2m) at (1.4,0)  { $\mathbb{F}_{2m+2}$};
\draw  (2, 0) -- (2.5,0); 
\draw (2.8,0) node {\scriptsize $\cdots $};
\draw  (3.0, 0) -- (3.5,0); 
\draw (4,0)  node { $\mathbb{F}_{6}$};
\draw (4.5,0) -- (5.5,0);
\draw (6,0) node { $\mathbb{F}_{4}$};
\draw (6.5,0) -- (7.5,0);
\draw (8,0) node { $\mathbb{F}_0^{2}$};
\draw (8.5,0) -- (9.5,0);
\draw (10,0) node { $\mathbb{F}_0$};
\end{tikzpicture}\,,
\ee
which results in the gauge theory duality
\be
SU(m+3)_{2m} + 2 \Lambda^2  \quad \longleftrightarrow\quad  Sp(m+1)_{(m+1) \pi}  -SU(2)_{(m+1)\pi}
\ee
For $m$ odd, both sides have 1-form symmetry  $\Gamma=\mathbb{Z}_2$.

\paragraph{Duality (3.80)} 
With the CS level\footnote{Note that earlier versions of \cite{Bhardwaj:2019ngx} had a typo in (3.79), which made the duality inconsistent with the 1-form symmetry. This was subsequently corrected.}
\be
k_{m,n,p} = p-m + A_{m+1, n+m+p+1} + A_{m+n, m+n+p+1} \,,
\ee
where  $A_{m,n}$ is the anomaly coefficient from integrating out $\Lambda^m$ for $SU(n)$ (see \cite{Aranda:2009wh}) 
\be
A_{m,n}= \frac{(n-2 m) \Gamma (n-2)}{\Gamma (m) \Gamma (n-m)}\,,\quad m\geq 3\,,\qquad A_{2,n}= n-4\,,\qquad A_{1,n}= 1\,.
\ee
Then there is a duality
\be
SU(m+p+4)_{k_{m,3,p}} + 2 \Lambda^{m+1} + 2 \Lambda^{m+3}  \quad \longleftrightarrow \quad 
Sp(m+1)_{(m+1)\pi} - SU(2)_{(m+p)\pi} - Sp (p+1)_{(p+1)\pi} \,.
\ee
For $m, p$ both odd this has a $\Gamma = \mathbb{Z}_2$  1-form symmetry. 
The surface configuration for example for $m=p=1$ is 
\be
\mathbb{F}_4  \qline \mathbb{F}_0^2 
\qline \mathbb{F}_0 \qline \mathbb{F}_0^2  \qline \mathbb{F}_4 \,.
\ee

\paragraph{Duality (3.178)} 
\be
\text{Spin}(2m+6) + \Lambda^{m+1} + 2 S   \quad \longleftrightarrow\quad  \Lambda^2 -
SU(m+3)_{{5\over 2} (m-1)}  \stackrel{\Lambda^2 \qquad }{\line(1,0){20}}  SU(2)_\pi\,.
\ee
This duality has a $\mathbb{Z}_2$ 1-form symmetry for $m$ odd, which can be confirmed from the geometry 
\be
\begin{tikzpicture}
\node (F2m) at (1.7,0)  { $\mathbb{F}_{2m}$};
\draw  (2, 0) -- (2.5,0); 
\draw (2.8,0) node {\scriptsize $\cdots $};
\draw  (3.0, 0) -- (3.5,0); 
\draw (4,0)  node { $\mathbb{F}_{4}$};
\draw (4.5,0) -- (5.5,0);
\draw (6,0) node { $\mathbb{F}_{2}$};
\draw (6.5,0) -- (7.5,0);
\draw (8,0) node { $\mathbb{F}_0^{2}$};
\draw (8.5,0) -- (9.5,0);
\draw (10,0) node { $\mathbb{F}_0^{1}$};
\draw (8.0,2)  node { $\mathbb{F}_{2}$};
\draw  (8, 0.3) -- (8,1.7); 
\end{tikzpicture}\,.
\ee

\paragraph{Duality (3.258)} 
\be
E_6 + \Lambda^{3} + 2 S   \quad \longleftrightarrow\quad  
\Lambda^3 - SU(6)_{0}  \stackrel{\Lambda^3 \quad }{\line(1,0){20}}  SU(2)_\pi\,.
\ee
This duality has a $\mathbb{Z}_3$ 1-form symmetry on both sides, as $S$ is uncharged under the center.

\paragraph{Duality (3.285)} 
\be
E_7 + \Lambda^{4} + C + C^2   \quad \longleftrightarrow\quad  
\Lambda^2 - SU(6)_{6}  \stackrel{\Lambda^2 \quad }{\line(1,0){20}}  SU(3)_{-19/2}\,.
\ee
This duality has a $\mathbb{Z}_2$ 1-form symmetry on both sides, as $C$ is uncharged under the center.

\paragraph{Duality (3.296)} 
\be
E_7 + \Lambda^{4} + 4C  \quad \longleftrightarrow\quad  
\Lambda^2 - SU(6)_{5}  \stackrel{\Lambda^2 \quad }{\line(1,0){20}}  SU(2)_{0} - SU(2)_\pi\,.
\ee
This duality has a $\mathbb{Z}_2$ 1-form symmetry on both sides.

\paragraph{Duality (3.300)} 
\be
E_7 + 4C + C^2  \quad \longleftrightarrow\quad  
3S  - \text{Spin}(12)  \stackrel{S \quad }{\line(1,0){20}}  SU(2)_\pi\,.
\ee
This duality has a $\mathbb{Z}_2$ 1-form symmetry on both sides. 

There are in addition numerous dualities with $Spin(N)$ and vector matter, which exhibit non-trivial one-form symmetries.

\section{5d Gauge Theories from Toric Calabi-Yau Threefolds}
\label{sec:Toric}

\subsection{1-Form Symmetry from Toric Data}

An interesting class of examples of Calabi-Yau singularities comes from toric geometry (see \cite{Closset:2018bjz, Eckhard:2020jyr} for a recent discussion in the context of 5d SCFTs).  These can be constructed from toric diagrams, which are sets of points, $v_i \in \Z^3$, $i=1,...,n$, whose hull forms a convex polytope.  The Calabi-Yau condition forces them to lie in a plane, which we may take to be at $v_3=1$, and so write $v_i=(w_i,1)$, where $w_i \in \Z^2$.  In general there may be internal points, which correspond to compact divisors, and external points, which correspond to non-compact divisors.  We label these respectively as
\be  \hat{v}_i, \;\;\; i=1,...,n_I, \;\;\;\;\; v_i, \;\;\; i=1,...,n_E \;, \ee 
where $n=n_I+n_E$.   The compact divisors associated to the internal points give rise to the gauge group generators, and the non-compact divisors associated to the external points give rise to flavor symmetry generators (after quotienting by some relations).  Explicitly, we find:
\be n_I = r, \;\;\;\; n_E = f+3 \;, \ee
where $r$ and $f$ are the ranks of the gauge and flavor groups.  Edges correspond to cycles, and taking a maximal triangulation gives rise to a complete resolution of the singularity, with internal edges giving the resulting exceptional curves.

The cohomology groups can be straightforwardly computed from this data.  For example, the group $H_4^{\partial X}(X)$ is generated by all compact and non-compact divisors.  If we define the $n \times 3$ matrix
\be A = \begin{pmatrix} \; A_I  \\ \text{\textemdash}  \\   A_E \; \end{pmatrix}\,, \ee
where
\be A_I = \begin{pmatrix} \hat{v}_1 \\ \vdots \\  \hat{v}_{n_I} \end{pmatrix} \;, \;\;\;\; A_E = \begin{pmatrix} v_1 \\ \vdots \\ v_{n_E} \end{pmatrix} \,,\ee
then we have
\be H_4^{\partial X}(X)  \; \cong \; \Z^n/\text{im}(A)\,.
\ee
To compute the (compact) homology group, $H_4(X)$, we note that this is generated by the internal points, of the form
\be \label{comtor} \begin{pmatrix} x_I \\ \text{\textemdash} \\ 0 \end{pmatrix}  \in \Z^n \;,\ee
which generate a $\Z^{n_I}$ subgroup of $\Z^n$.  The divisors which are equivalent to these are ones related by an element in $\text{im}(A)$.  Let us denote this group by $\Z^{n_I} + \text{im}(A)$.  Then,
\be H_4(X) = (\Z^{n_I} + \text{im}(A))/\text{im}(A) \cong \Z^{n_I}/(\Z^{n_I} \cap \text{im}(A)) \cong \Z^{n_I} \;, \ee
where in the second equality we used the second isomorphism theorem, and in the third we observe that $\Z^{n_I} \cap \text{im}(A)$ is trivial, since $A_E$ has trivial kernel.\footnote{Namely, any valid toric diagram has at least 3 external points which are linearly independent over $\Z^3$, so there is no element in the image of $A$ of the form \eqref{comtor}.} 

From \eqref{h4gammadef}, the $1$-form symmetry group and flavor rank are given by
\be \label{gammator2} H_4^\partial(X)/f_4(H_4(X))  \cong \Z^f \oplus \Gamma \;,
\ee
where $f_4$ is defined in \eqref{h2les}.  In the present case, $H_4(X)$ is a subgroup of $H_4^\partial(X)$, and this map is simply the inclusion map. Then we have, using the third isomorphism theorem
\be H_4^\partial(X)/f_4(H_4(X))  \cong ( \Z^n/\text{im}(A) ) /  \left((\Z^{n_I} + \text{im}(A))/\text{im}(A) \right) \cong \Z^n/(\Z^{n_I} \oplus \text{im}(A))  \,.\ee
Note that both $\Z^n \cong \Z^{n_I} \times \Z^{n_E}$ and $\Z^{n_I} + \text{im}(A) \cong \Z^{n_I} \times \text{im}(A_E)$ are product groups with a common factor, and so their quotient is simply given by
\be H_4^\partial(X)/f_4(H_4(X)) \cong \Z^{n_E}/\text{im}(A_E)\,. \ee
In particular, note that this is independent of the internal points, $A_I$.

We may compute this quotient by writing $A_E$ in Smith normal form as
\be A_E = SDT \;, \ee
where $D$ is diagonal, explicitly,
\be D = \begin{pmatrix} 
	\alpha_1 & 0 & 0\\
	0 & \alpha_2 & 0\\
	0 & 0  & \alpha_3 \\
	0 & 0 & 0 \\ 
	& \vdots & \end{pmatrix} \,, 
\ee
with $n_E-3 = f$ zero rows {(and $\alpha_1 | \alpha_2 | \alpha_3$)}.  Then, from \eqref{gammator2}, we find the 1-form symmetry group is
\be \label{gammatoric} \Gamma=\bigoplus_{i=1}^3\Z/\alpha_i \Z \;, \ee
and we see that $f$ is the rank of the flavor symmetry, as expected.

Note that this prescription is manifestly independent of the resolution, $\bX$ (which would correspond to a choice of triangulation of the toric diagram).  One way to see this is to observe that, for any choice of triangulation, we have:
\be H_2(X) \cong H_\partial^4(X) \cong  \text{Hom}(H_4^\partial(X),\Z) \cong \text{ker}(A^T) \ee
The different choices of triangulation simply identify different curves with different generators of $\text{ker}(A^T)$. As long as the external vertices remain the same, the choice of triangulation of the convex region they enclose will not change the quotient, and so neither the 1-form symmetry.

\subsection{Examples}

\begin{figure}
	\centering
	\includegraphics*[width=8cm]{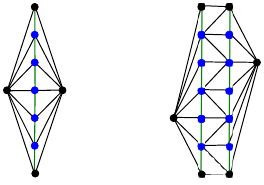}
	\caption{The toric diagrams for $SU(N)_k$ and $SU(N)_{\frac{N}{2} + k_1} \times SU(N)_{\frac{N}{2} + k_2}$ linear quiver. On the left the case of $k=0$ for $N=6$ is shown, on the right hand side $k_1= 1=k_2$. In both diagrams we show a  specific full triangulation, \ie, the geometry in question is the fully resolved one, \ie, the theory on the Coulomb branch. The vertical green lines indicate the rulings that give the $SU(N)$ gauge groups. \label{fig:SUNToric}}
\end{figure}

The first class of examples we consider are the pure SYM theories, which were discussed in section \ref{sec:SYM}. For instance, the toric diagram for $SU(N)_k$ is shown in figure  \ref{fig:SUNToric}. The 1-form symmetry was determined to be $\Gamma = \mathbb{Z}_{\text{gcd}(N,k)}$.

\begin{figure}
\centering
\includegraphics[width= 10cm]{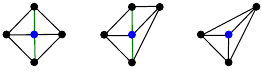}
\caption{Toric diagrams for $SU(2)_0$, $SU(2)_\pi$ and $\mathbb{P}^2$ theory (which is obtained from $SU(2)_\pi$ by an external flop). The internal vertex (blue) indicates the Cartan of the rank one  gauge group, the external vertices correspond to non-compact divisors. Whenever there is a ruling, i.e. a partial singularization to an $A_1$ singularity, this is indicated by a green line. 
\label{fig:SU2T}}
\end{figure}
The simplest examples are the rank 1 theories \cite{Seiberg:1996bd}: $SU(2)_0$, where the toric diagram is a diamond, $\{v_i \} = \{ (\pm 1,0),(0,\pm 1) \}$, shown in figure \ref{fig:SU2T}. This theory has flavor rank $f=1$ (coming from the instantonic symmetry, which is enhanced to $SU(2)$ in this case), and $\Gamma=\Z_2$, which is indeed the expected $1$-form symmetry for this gauge theory.
The theory $SU(2)_\pi$ has the toric diagram modified by moving $(1,0) \rightarrow (1,1)$, see figure \ref{fig:SU2T}..  Then we again find $f=1$, which is a $U(1)$, but now $\Gamma=1$, \ie, there is no $1$-form symmetry. This is consistent also with our observations on $Sp(N)$ theories and the dependence of the 1-form symmetry on the theta angle, as discussed in section \ref{sec:SYM}. All theories $SU(2) + N_F \bm{F}$ with $N_F=1, \cdots ,7$ (which are mass deformations of SCFTs)  do not have any 1-form symmetry due to the fundamental matter.  For $N_F=1,2$, these theories have a toric geometric description, and this fact again is confirmed from the geometry. 

At rank 1 there is one other theory, which has a 1-form symmetry, which is the non-Lagrangian theory that is obtained by blowing up
$\C^3/\Z_3$ (local $\mathbb{P}^2$). Alternatively it is obtained by a flop from the $SU(2)_\pi$ theory. Here the toric diagram has three external points, $w_i \in \{ (-1,0),(0,-1),(1,1)\}$, and one internal point, $w_i =(0,0)$ as shown in figure \ref{fig:SU2T}.
We find $f=0$ and 
\be
\mathbb{P}^2:\qquad \Gamma=\Z_3\,.\ee  
This theory has no flavor symmetry, but a non-trivial $1$-form symmetry.

As another, more complicated example, consider a toric diagram with external points at
\be\label{SUNk1k2}
w_i \in \left\{ \; (0,0),(0,N),(1,0), (1,N),\left(-1,\frac{N}{2}+k_1\right), \left(2,\frac{N}{2}-k_2\right) \; \right\}  \,, \ee
shown in figure \ref{fig:SUNToric}.  
This engineers an $SU(N)_{\frac{N}{2} + k_1} \times SU(N)_{\frac{N}{2} + k_2}$ linear quiver theory.  The $1$-form symmetry of this theory can be computed as follows: before considering the matter and CS terms, it is $\Z_N \times \Z_N$ from the two $SU(N)$  factors.  The bifundamental matter then breaks this to the diagonal $\Z_N$ subgroup, and the two CS terms finally break it to:\footnote{To see that it is the shifted CS level, $k_i$, that is relevant here, we recall, following section \ref{sec:5dhfs}, that after we integrate out the bifundamental matter, we are left with pure CS theories with levels $k_i$ or $k_i + N$, depending on the sign of the mass.  For either choice, we find the 1-form symmetry above. 
} 
\be \label{gammatq}\Gamma = \Z_{\text{gcd}(N,k_1,k_2)} \,.
\ee
This is in agreement with the 1-form symmetry computed from the toric diagram. 
The quivers above appear to be the most general that are (i) strictly convex, (ii) give rise to an ``allowed vertical reduction,'' in  the notation of \cite{Closset:2018bjz}, leading to a standard gauge theory description, and (iii) have non-trivial $1$-form symmetry. 

Finally in the case $N=2$, recall that there is not a proper CS term, but only a discrete theta angle.  Roughly speaking, we may identify the theta angle with $\theta=k \pi \; (\text{mod} \; \pi)$.  As discussed in section \ref{sec:5dhfs}, the $\Z_2$ 1-form symmetry of an $SU(2)$ theory is broken for $\theta=\pi$, so \eqref{gammatq} will continue  to hold in this case.  For example, for the above quiver in the case $N=2$, $k_1=k_2=1$, which is referred to as the ``beetle quiver,'' \cite{Closset:2018bjz}, which is also a descendant of the $(D_5, D_5)$ conformal matter theory \cite{Jefferson:2018irk, Apruzzi:2019opn}, we find $\theta=\pi$ for the two gauge groups, which breaks the putative 1-form symmetry.  This is consistent with the fact that this theory has a dual description as an $SU(3)$ theory with $N_f=2$, which  has no $1$-form symmetry.

\subsection{$T_N$ and Descendants}

Another class of toric Calabi-Yau threefolds corresponding to 5d gauge theories with interacting UV fixed points are the theories $T_N$. They are constructed as M-theory on the singularity $\mathbb{C}^{3}/(\mathbb{Z}_N\times \mathbb{Z}_N)$. They have a description in terms of toric geometry and (dual) brane-webs \cite{Benini:2009gi}.  More recently, their descendants, that are  obtained by decoupling hypermultiplets, were studied, and their flavor symmetries and Higgs branches were determined \cite{Eckhard:2020jyr}. In particular it was shown that there is a large class of non-Lagrangian descendants, which turn out to have non-trivial 1-form symmetries (as well as 0-form symmetries).

\begin{figure}
	\centering
\subfloat[]{\includegraphics[width=5cm]{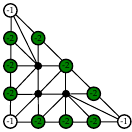} }
\subfloat[]{	\includegraphics*[width=5cm]{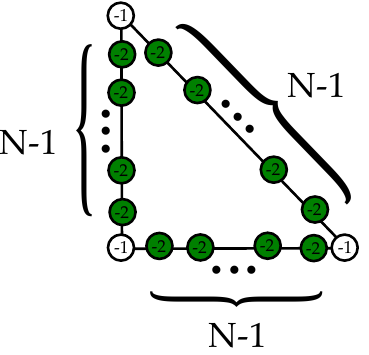}}
	\caption{(a) Fully triangulated toric fan for $T_4$: the internal vertices are the compact divisors, associated to the Cartans of the gauge group. We marked the external vertices by the self-intersection numbers of the non-compact divisors with the compact divisors. These form the 	combined fiber diagram (CFD): $(-2)$ are the marked vertices, which form the subgraph that is the Dynkin diagram of the flavor symmetry $G_F= SU(N)^3$.  $(-1)$ are the curves with self-intersection number $-1$, which correspond to matter hypermultiplets in bifundamentals in pair-wise combinations of the $SU(N)^3$ flavor symmetry. (b) the CFD for $T_N$. \label{fig:CFDTN}}
\end{figure}

The $T_N$ theories are rank $r=(N-1)(N-2)/2$ theories, and have flavor symmetry $G_F= SU(N)^3$ (which for $N=3$ enhances to $E_6$ -- the rank 1 Seiberg theory, that is the strong coupling limit of $Sp(1) + 5 \bm{F}$). 
The toric fan for the singular model, \ie, the SCFT, is 
\be
T_N:\qquad A_E = ((0,0,1), (N,0,1), (0,N,1)) \,.
\ee
Applying the toric formalism to this, it is easy  to check using  \eqref{gammatoric} that there is no 1-form symmetry for any of the $T_N$ theories. 
A field theory way to see this is to recall that there is a linear quiver description of $T_N$ as 
\be
[N]- SU(N-1) - SU(N-2) - \cdots - SU(2) -[2] \,,
\ee
where we see the 1-form symmetry is completely broken by the fundamental and bifundamental matter.

\begin{figure}
	\centering
	\includegraphics*[width=15cm]{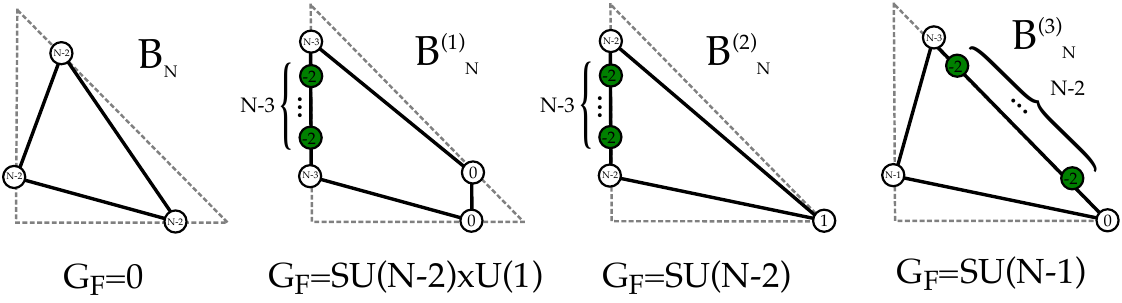}
	\caption{The non-Lagrangian theories determined in \cite{Eckhard:2020jyr} obtained from $T_N$ after decoupling all hypermultiplets, as well as their flavor symmetries $G_F$.\label{tab:Bns}}
\end{figure}

However, once flavors are decoupled the theories can have higher-form symmetries. 
For instance, one of the descendants of $T_3$ is the theory with geometry $\mathbb{P}^2$, which has $\mathbb{Z}_3$ one-form symmetry. 
Similarly for $T_N$ the endpoints of the decoupling tree (where we successively decouple hypermultiplets) has a non-trivial one-form symmetry: these theories are denoted by $B_N$ (as in ``bottom of $T_N$'') \cite{Eckhard:2020jyr}, and like the rank one case, do not admit an IR description in terms of a non-abelian gauge theory.\footnote{This can either be seen from the toric description, where there is no consistent ruling, that results in a weakly-coupled gauge theory. Alternatively, we can apply the criterion using BG-CFDs (box graph CFDs), which are IR-versions of CFDs \cite{Hayashi:2014kca, Apruzzi:2019enx}. It is necessary for a gauge theory description to exist, that its BG-CFD can be embedded into the CFD, which in these instances is not possible. }
Since these models are toric, we can apply the general formalism above. 
These models have external vertices, which are precisely the vertices of the CFD of these theories, given by
\be
B_N:\qquad A_E= ((N - 1, 0, 1), (1, N- 1, 1), (0, 1, 1)) \,,
\ee
where each is a curve of self-intersection $N-2$ \cite{Eckhard:2020jyr}. The CFD is shown in figure \ref{fig:CFDTN}.
The Smith normal form results in the above 1-form symmetry $\Gamma$:
\be
\begin{array}{c|c|c|c}
	B_N & \text{Gauge Rank} & \text{Flavor Rank} & \Gamma \cr \hline \hline 
	B_3 = \mathbb{P}^2 & 1 & 0 & \mathbb{Z}_3 \cr \hline 
	B_4 & 3 & 0 & \mathbb{Z}_7 \cr \hline 
	B_5 & 5 & 0 & \mathbb{Z}_{13} \cr \hline 
	B_6 & 10 & 0 & \mathbb{Z}_{21}\cr \hline 
	B_7 & 15 & 0 & \mathbb{Z}_{31} \cr \hline \hline
	B_N & {1\over 2} (N-1)(N-2) & 0 & \mathbb{Z}_{N(N-3) +3} \cr 
\end{array}
\ee

These are generically not the only theories that are descendants of $T_N$ and have a non-trivial 1-form symmetry. 
For rank 1 we know already the theory $SU(2)_0$ which has $\Gamma =\mathbb{Z}_2$. 

For $T_4$ there are three more descendants with non-trivial  1-form symmetry. The matrix $A_E$ is obtained from the CFD, simply by recalling which coordinates the vertex has in the initial toric diagram figure \ref{fig:CFDTN}.

\begin{table}
	$$
	\begin{array}{c|c|c|c}
	A_E & \text{Toric Fan/CFD} & G_F& \Gamma  \cr \hline 
	\left(
	\begin{array}{ccccc}
	4 & 3 & 2 & 1 & 0 \\
	0 & 1 & 2 & 3 & 1 \\
	1 & 1 & 1 & 1 & 1 \\
	\end{array}
	\right) 
	& \includegraphics[width=4cm]{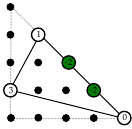} & SU(3) & \mathbb{Z}_3 \cr \hline 
	\left(
	\begin{array}{ccccc}
	3 & 3 & 0 & 0 & 0 \\
	0 & 1 & 3 & 2 & 1 \\
	1 & 1 & 1 & 1 & 1 \\
	\end{array}
	\right)
	& \includegraphics[width=4cm]{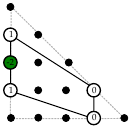} & SU(2)\times U(1) & \mathbb{Z}_3 \cr \hline 
	\left(
	\begin{array}{cccc}
	4 & 0 & 0 & 0 \\
	0 & 3 & 2 & 1 \\
	1 & 1 & 1 & 1 \\
	\end{array}
	\right) 
	& \includegraphics[width=4cm]{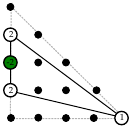} & SU(2) & \mathbb{Z}_4 \cr \hline 
	\left(
	\begin{array}{ccc}
	3 & 1 & 0 \\
	0 & 3 & 1 \\
	1 & 1 & 1 \\
	\end{array}
	\right)
	& \includegraphics[width=4cm]{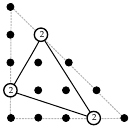} &\emptyset & \mathbb{Z}_7 \cr \hline 
	\end{array}
	$$
	\caption{Descendants of the $T_4$ theory, which have non-trivial 1-form symmetry. First we give the toric vectors for the vertices of the curves in the toric diagram and its boundary, which is the CFD, that is shown in the second column. The flavor symmetry is indicated by $G_F$ and $\Gamma$ is the 1-form symmetry. 
		\label{tab:T41s}}
\end{table}

\begin{table}
	$$
	\begin{array}{c|c|c|c}
	A_E & \text{Toric Fan/CFD} & G_F& \Gamma  \cr \hline 
	\left(
	\begin{array}{ccccccc}
	6&5 & 4 & 3 & 2 & 1&0 \\
	0 & 1 & 2 & 3&4 & 1 & 1\\
	1 & 1 & 1 & 1 & 1&1 &1  \\
	\end{array}
	\right) &
	\includegraphics[width=4cm]{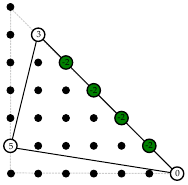} & SU(5) & \mathbb{Z}_5 \cr \hline 
	\left(
	\begin{array}{ccccccc}
	5&5 & 0 & 0 & 0 & 0&0 \\
	0 & 1 & 5 & 4&3 &2 & 1\\
	1 & 1 & 1 & 1 & 1&1 &1  \\
	\end{array}
	\right) &
	\includegraphics[width=4cm]{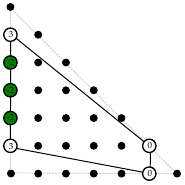} & SU(4)\times U(1) & \mathbb{Z}_5 \cr \hline 
	\left(
	\begin{array}{cccccc}
	6& 0 & 0 & 0 & 0&0 \\
	0 &  5 & 4 & 3&2 & 1 \\
	1 & 1 & 1 & 1 & 1&1   \\
	\end{array}
	\right) &
	\includegraphics[width=4cm]{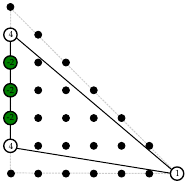} & SU(4) & \mathbb{Z}_6 \cr \hline 
	\left(
	\begin{array}{ccc}
	5 & 1 & 0 \\
	0 & 5 & 1 \\
	1 & 1 & 1 \\
	\end{array}
	\right)&
	\includegraphics[width=4cm]{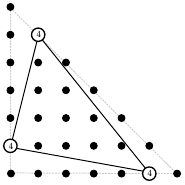} &\emptyset & \mathbb{Z}_{21} \cr \hline 
	\end{array}
	$$
	\caption{Descendants of the $T_6$ theory, which have non-trivial 1-form symmetry. First we give  the CFD, flavor symmetry is indicated by $G_F$ and $\Gamma$ is the 1-form symmetry. 
		\label{tab:T61s}}
\end{table}

Here we have only shown one toric triangulation as well as the CFD, which are the curves that are the intersections of compact with non-compact divisors. 
The descendants are obtained by toric flops, or more efficiently, CFD-transitions. The models with non-trivial 1-form symmetry are shown in tables \ref{tab:T41s} and \ref{tab:T61s}.
The general case is shown in figure \ref{tab:Bns}.  Computing the 1-form symmetry, we find for these non-Lagrangian models ($N>2$)
\be
\begin{array}{c|c|c}
	\text{Theory} &  \text{Flavor Symmetry} & \Gamma \cr \hline \hline 
	B_N & 0 & \mathbb{Z}_{N(N-3) +3} \cr \hline
	B^{(1)}_N & SU(N-2)\times U(1) & \mathbb{Z}_{N-1} \cr \hline
	B^{(2)}_N & SU(N-2) & \mathbb{Z}_{N} \cr \hline
	B^{(3)}_N & SU(N-1) & \mathbb{Z}_{N-1} \,.
\end{array}
\ee
It would indeed be interesting to find alternative methods to confirm these higher-form symmetries.

\section{Higher-Form Symmetries from 6d to 5d}
\label{sec:6dto5d}

In this section we investigate the relation between the higher-form symmetries of 5d theories and their 6d parent theories. All known 5d SCFTs are obtained by circle-reduction from 6d with additional mass deformation (and possibly the action of automorphisms).  To study this {question}, we will use F-theory realizations of 6d theories in  terms of elliptically fibered Calabi-Yau threefolds, and use the F-theory/M-theory duality to realize their 5d compactifications as M-theory on the same elliptic Calabi-Yau.  We saw a simple example of this in section \ref{sec:n2ex}.

One interesting aspect of this correspondence is that the 6d theories themselves may have higher-form symmetries, which we can relate to those of their 5d reductions.  In particular, many 6d theories are not genuine QFTs, but instead only defined as so-called ``relative'' 6d theories  \cite{Witten:2009at,Freed:2012bs}.  Instead of a partition function, they have a partition vector, analogous to the set of conformal blocks of a 2d CFT.  This data can be characterized by a finite abelian group assigned to each 6d theory, called the ``defect group'' \cite{DelZotto:2015isa}, which  can be thought of as a ``self-dual'' 2-form symmetry \cite{Garcia-Etxebarria:2019cnb,Eckhard:2019jgg}.  We briefly review this structure below.  Upon dimensional reduction to 5d, this gives rise to a 1-form symmetry.\footnote{Equivalently, after gauging, we may instead obtain a 2-form symmetry in 5d.  However, we do not obtain both a 1-form and 2-form symmetry, as would  usually be the case in such a reduction, precisely because of the self-dual nature  of the 2-form symmetry \cite{Eckhard:2019jgg}.}  In addition, the 6d theory may itself have a 1-form symmetry, in which  case this can also contribute to the 5d 1-form symmetry.  Schematically, we have
\be \label{5d6d1form}
\Gamma^{1-\text{form}}_{5d} = \Gamma^{1-\text{form}}_{6d} \oplus \Gamma^{\text{defect}}_{6d} \,.
\ee
Below we will demonstrate that the 1-form symmetry on the LHS can be read off from the topology of the elliptic Calabi-Yau of the 6d F-theory model according to the discussion of section \ref{sec:higherformmtheory}.

The circle-reduction of theories with non-trivial defect group, which are so-called non-very Higgsable theories in 6d, gives rise to 5d theories with non-trivial 1-form symmetry. In general a 6d theory will be a combination of minimal conformal matter theories (which typically have matter content which breaks the 1-form symmetry in 6d), and a collection of curves with possibly tuned gauged groups. These can give rise to 1-form symmetries in 6d, if the matter is invariant under the action of the center of the total gauge group. We will encounter examples of this type.

\subsection{Defect Group in 6d}

As was first understood for the 6d $\cN=(2,0)$ theories \cite{Witten:2009at,Freed:2012bs,Tachikawa:2013hya}, and later for more general 6d $\cN=(1,0)$ theories \cite{DelZotto:2015isa}, many 6d theories are not genuine QFTs, but have some subtle properties which make them more analogous to chiral CFTs in 2d. For example, rather than having a unique partition function on a closed spacetime manifold, $M_6$, they have a ``partition vector,'' similar to a set of conformal blocks.  The elements of the partition vector are labeled by a (non-canonically) chosen Lagrangian subgroup of  $H^3(M_6,\CC)$, where the coefficient group, $\CC$, is a finite abelian group known as the ``defect group.''   This is a similar structure to a 2-form symmetry with group $\Gamma$, whose observables would be labeled by the entire group $H^3(M_6,\Gamma)$.  For this reason we may refer to this structure as a ``self-dual'' 2-form symmetry \cite{Garcia-Etxebarria:2019cnb,Eckhard:2019jgg}.  Upon dimensional reduction to 5d, this  descends to either a 1-form or 2-form symmetry, depending on the choice of  Lagrangian subgroup.

Six-dimensional $\cN=(1,0)$ theories have a realization in terms of F-theory on a singular elliptic Calabi-Yau threefold, 
\be
\mathbb{E} \hookrightarrow  M \rightarrow B_2 \,,
\ee
where the base is 
\be
B_2 = \mathbb{C}^2/\Gamma_B\,,\qquad \Gamma_B  \subset U(2) \,.
\ee
The defect group $\mathcal{C}$ of the associated theory was determined in \cite{DelZotto:2015isa} as the abelianization of $\Gamma_B$ 
\be
\mathcal{C} = \text{Ab}[\Gamma_B] \,.
\ee
The defect groups in 6d were classified and have the following types 
\be
\begin{array}{c||c|c|c|c|c|c}
\text{Theory Type} & A   & D^{(\text{even})} & D^{(\text{odd})} & E_6 & E_7 & E_8  \cr \hline
\mathcal{C}  & \mathbb{Z}_{k+1} &\mathbb{Z}_{2} \times \mathbb{Z}_{2k} & \mathbb{Z}_{4k} &  \mathbb{Z}_3 & \mathbb{Z}_2 &1 \cr 
\end{array}
\ee
Here $A$ refers to a linear quiver of curves 
\be
A:\qquad n_1 \cdots n_m 
\ee
in the tensor branch of the geometry, where the curves have self-intersections $-n_j$.
The defect group is $\mathbb{Z}_p$ where $p/q$ is the continued fraction of $n_j$.

The D-type quivers are curves in the base arranged in shape of a D-type Dynkin diagram with negative self-intersection numbers
\be
D:\qquad 
 2 \stackrel{2}{n}  n_1 \cdots, n_m \,.
\ee
For D-type the possible groups that can occur in 6d are limited to 
\be
D^{(\text{even})}\ :\qquad \mathbb{Z}_4\,,\ \mathbb{Z}_8 \,,\ \mathbb{Z}_{12} \,,\qquad 
D^{(\text{odd})}\ :\qquad \mathbb{Z}_2\times \mathbb{Z}_\ell \,,\ \ell= 2,4,6 \,.
\ee
Finally, $E$ type corresponds to $-2$ curves arranged in the Dynkin diagram of the associated Lie algebra. 
For the 6d $(2,0)$ theories of ADE type, the defect group  is $\mathcal{C} ={Z}(\mathfrak{g}_{ADE})$, \ie, the center of the associated ADE algebras.
If the defect group in 6d is non-trivial, the theories are called non-very Higgsable, and we will see that the 5d theory inherits the defect group as part of its 1-form symmetry.

\subsection{NHCs}

The non-Higgsable clusters (NHCs) \cite{Morrison:2012np} are essential building blocks for 6d SCFTs and examples of non-very Higgsable theories. The single node NHCs have a  tensor branch description as 
\be\label{NHC}
\stackrel{\mathfrak{g}}{n} \,,
\ee
where the base is a single $-n$ self-intersection curve and $\mathfrak{g}$ is the non-Higgsable gauge algebra
\be
\begin{tabular}{c||c|c|c|c|c|c|c|c}
$-n$ & $-3$& $-4$& $-5$& $-6$& $-7$& $-8$ &$-12$ \cr \hline
Gauge algebra $\mathfrak{g}$ & $\mathfrak{su}(3)$ &$\mathfrak{so}(8)$ &$\mathfrak{f}_4$ &$\mathfrak{e}_6$ &$\mathfrak{e}_7+ {1\over 2} \bm{56}$ &$\mathfrak{e}_7$ &$\mathfrak{e}_8$ 
\end{tabular}
\ee
The base of the F-theory model is an orbifold by 
\be
\Gamma_B :\ (z, w) \sim (\omega z, \omega w )\,,\qquad \omega = e^{2 \pi i/n} \,.
\ee
The base is of A-type and contributes a defect group $\mathcal{C}= \mathbb{Z}_n$.  In addition, there is a 1-form symmetry in 6d, given by the center, $Z_G$, of the simply connected gauge group with algebra $\mathfrak{g}$ (except for $n=7$, where we instead have  trivial 1-form symmetry due to the matter).  

From \eqref{5d6d1form}, we then expect that 1-form symmetry of the 5d theory is given by
\be\label{OneFormNHC}
\Gamma = \mathbb{Z}_n \oplus {Z}_{G} 
\ee
for the NHC (\ref{NHC}).  To reproduce this using the 5d description, note that in the dimensional reduction, it is observed in \cite{DelZotto:2017pti,  Apruzzi:2019kgb} that -- after decoupling of a single compact surface --  there is a  pure  $G$ gauge theory without matter, except for the $-7$ case which is an $E_7$ with ${1\over 2} {\bf 56}$. 
For the SCFT sector, we then expect that the center of the NHC group leads to a $Z_G$ 1-form symmetry in 5d.  

However, we also expect a $\Z_n$ 1-form symmetry, which we claim comes from the decoupled sector, as follows.  First consider starting with the pure $\cN=1$ $G$ gauge theory in 5d, which has a $U(1)_{\text{inst}}$ symmetry.  We may then gauge this symmetry, which amounts to including a new $U(1)$ gauge multiplet with a topological coupling involving the instanton density of the $G$ gauge field. Specifically, we suppose that we gauge this symmetry with charge $n$, \ie, we include $n$ times the minimal topological coupling.  Then we claim this theory provides a 5d description of the NHC associated to a $G$ gauge group on a $-n$ curve.  Namely, as we describe below, the geometry of the NHC is given by a configuration of intersecting (blown up) Hirzebruch surfaces associated to the affine Dynkin diagram of the associated group, as shown in table \ref{tab:NHCS}.  Correspondingly, we may define the prepotential of the theory on the extended Coulomb branch in terms of parameters, $\varphi_i$ , $i=0,...,r_G$ associated to the Dynkin nodes.   Then one checks that if we identify $\varphi_0$ with the Coulomb parameter of the $U(1)$ gauge field and $\varphi_i$,  $i=1,...,r_G$ with those of the $G$ gauge field in the theory above, and make the redefinitions 
\be  \varphi_i  \; \rightarrow \;  \varphi_i + d_i \varphi_0 \,,
\ee
where $d_i$ are the Dynkin indices, then the prepotential of this gauge theory precisely agrees with that computed from the geometry of the NHC.  We may then take a limit where  the gauge coupling of the $U(1)$ goes to infinity, which decouples the gauge field, and leaves the pure $G$ gauge theory, as above.  The upshot  of this  interpretation is that, in addition to the ${Z}_G$ 1-form symmetry from the $G$ gauge field, the $U(1)$ gauge field also contributes a 1-form symmetry.  Due to the coupling to the instanton symmetry with charge $n$, the $U(1)$ 1-form symmetry of the $U(1)$ gauge field is broken to a $\Z_n$ subgroup, which is precisely the contribution to the  5d 1-form symmetry from the defect group in 6d.

We can now compare this to the result from studying the NHC  geometries and applying the results of section \ref{sec:higherformmtheory}. 
For even $n$, the surfaces,  $\mathbb{F}_i$, are glued together in the form of an affine $\mathfrak{g}$ Dynkin diagram, as shown in table \ref{tab:NHCS}. The Hirzebruch surfaces intersect as 
\be
\mathbb{F}_{m} \cdot \mathbb{F}_{m-2} = C_{m, m-2} \,,\qquad  C^2_{m, m-2}|_{\mathbb{F}_m} = m \,,\qquad C^2_{m, m-2}|_{\mathbb{F}_{m-2}} = - (m-2) \,.
\ee
The intersection matrix of compact surfaces $\mathbb{F}_{n_i} = S_i$ and compact curves gives in Smith normal form the 1-form symmetries of the theory. This is summarized in table \ref{tab:NHCS} and agrees with the expression (\ref{OneFormNHC}).

An alternative approach, which applies in the cases of even $n$, is to note
that the total space of the singular elliptically fibered Calabi-Yau, can be presented as a quotient \cite{Heckman:2013pva}
\be X= (\C^2 \times T^2)/\Z_n \,.
\ee
Here $\Z_n$ acts as:
\be (z_1,z_2,w) \sim (\omega \; z_1, \omega \; z_2, \omega^{-2} \; w) \,, \ee
where $\omega=e^{2 \pi i/n}$, $z_i$ are the coordinates of $\C^2$, and $w$ is the coordinate on $T^2$.   Then we argue in appendix \ref{sec:nhcquotient} that studying the homology of the boundary of this quotient space also gives the expected result  for the 1-form symmetry.

\begin{table}
\centering
$$
\begin{array}{c|c|c}
(n, \mathfrak{g}) & \text{Surface Configuration} & \Gamma = \mathbb{Z}_n \oplus {Z}_{G}\cr \hline \hline
 (3, \mathfrak{su}(3)) & 
\begin{tikzpicture}
\draw (0,0) node {\scriptsize $\mathbb{F}_1$};
\draw (1,0) node {\scriptsize $\mathbb{F}_1$};
\draw (0,1)  node {\scriptsize $\mathbb{F}_1$};
\draw  (0.3, 0) -- (0.7,0); 
\draw  (0, 0.3) -- (0,0.7) ;
\draw (0.3, .7) -- (0.7,0.3) ;
\end{tikzpicture}
& \mathbb{Z}_3 \times \mathbb{Z}_3 
 \cr \hline
 (4, \mathfrak{so}(8)) & 
\begin{tikzpicture}
\draw (0,0) node {\scriptsize $\mathbb{F}_0$};
\draw (1,0) node {\scriptsize $\mathbb{F}_2$};
\draw (0,1)  node {\scriptsize $\mathbb{F}_2$};
\draw (-1,0) node {\scriptsize $\mathbb{F}_2$};
\draw (0,-1)  node {\scriptsize $\mathbb{F}_2$};
\draw  (0.3, 0) -- (0.7,0); 
\draw  (0, 0.3) -- (0,0.7) ;
\draw  (-0.3, 0) -- (-0.7,0); 
\draw  (0, -0.3) -- (0,-0.7) ;
\end{tikzpicture}
& \mathbb{Z}_4 \times \mathbb{Z}_2\times \mathbb{Z}_2 
\cr \hline
 (5, \mathfrak{f}_4) & 
\begin{tikzpicture}
\draw (0,0) node {\scriptsize $\mathbb{F}_3$};
\draw (1,0) node {\scriptsize $\mathbb{F}_1$};
\draw (2,0)  node {\scriptsize $\mathbb{F}_1$};
\draw (3,0) node {\scriptsize $\mathbb{F}_6$};
\draw (4,0)  node {\scriptsize $\mathbb{F}_8$};
\draw  (0.3, 0) -- (0.7,0); 
\draw  (1.3, 0) -- (1.7,0); 
\draw[->, double ]  (2.3, 0) -- (2.7,0); 
\draw  (3.3, 0) -- (3.7,0); 
\end{tikzpicture}
& \mathbb{Z}_5 \cr\hline
 (6, \mathfrak{e}_6) & 
\begin{tikzpicture}
\draw (0,0) node {\scriptsize $\mathbb{F}_4$};
\draw (1,0) node {\scriptsize $\mathbb{F}_3$};
\draw (2,0)  node {\scriptsize $\mathbb{F}_0$};
\draw (3,0) node {\scriptsize $\mathbb{F}_2$};
\draw (4,0)  node {\scriptsize $\mathbb{F}_4$};
\draw (2,1) node {\scriptsize $\mathbb{F}_2$};
\draw (2,2)  node {\scriptsize $\mathbb{F}_4$};
\draw  (0.3, 0) -- (0.7,0); 
\draw  (1.3, 0) -- (1.7,0); 
\draw  (2.3, 0) -- (2.7,0); 
\draw  (3.3, 0) -- (3.7,0); 
\draw  (2, 0.3) -- (2,0.7); 
\draw  (2, 1.3) -- (2,1.7); 
\end{tikzpicture}
& \mathbb{Z}_6\times \mathbb{Z}_3 \cr\hline
 \left(7, \mathfrak{e}_7 + {1\over 2} \bm{56}\right) & 
\begin{tikzpicture}
\draw (0,0) node {\scriptsize $\mathbb{F}_5$};
\draw (1,0) node {\scriptsize $\mathbb{F}_3$};
\draw (2,0)  node {\scriptsize $\mathbb{F}_1$};
\draw (3,0) node {\scriptsize $\mathbb{F}_1$};
\draw (4,0)  node {\scriptsize $\text{Bl}_1\mathbb{F}_3$};
\draw (5,0)  node {\scriptsize $\mathbb{F}_4$};
\draw (6,0)  node {\scriptsize $\text{Bl}_1\mathbb{F}_8$};
\draw (3,1) node {\scriptsize $\text{Bl}_1\mathbb{F}_3$};
\draw  (0.3, 0) -- (0.7,0); 
\draw  (1.3, 0) -- (1.7,0); 
\draw  (2.3, 0) -- (2.7,0); 
\draw  (3.3, 0) -- (3.6,0); 
\draw  (4.4, 0) -- (4.7,0); 
\draw  (5.3, 0) -- (5.6,0); 
\draw  (3, 0.3) -- (3,0.7); 
\end{tikzpicture}
& \mathbb{Z}_7 \cr\hline
(8, \mathfrak{e}_7) & 
\begin{tikzpicture}
\draw (0,0) node {\scriptsize $\mathbb{F}_6$};
\draw (1,0) node {\scriptsize $\mathbb{F}_4$};
\draw (2,0)  node {\scriptsize $\mathbb{F}_2$};
\draw (3,0) node {\scriptsize $\mathbb{F}_0$};
\draw (4,0)  node {\scriptsize $\mathbb{F}_2$};
\draw (5,0) node {\scriptsize $\mathbb{F}_4$};
\draw (6,0)  node {\scriptsize $\mathbb{F}_6$};
\draw (3,1) node {\scriptsize $\mathbb{F}_2$};
\draw  (0.3, 0) -- (0.7,0); 
\draw  (1.3, 0) -- (1.7,0); 
\draw  (2.3, 0) -- (2.7,0); 
\draw  (3.3, 0) -- (3.7,0); 
\draw  (4.3, 0) -- (4.7,0); 
\draw  (5.3, 0) -- (5.7,0); 
\draw  (3, 0.3) -- (3,0.7); 
\end{tikzpicture}
& \mathbb{Z}_8\times \mathbb{Z}_2 \cr\hline
(12, \mathfrak{e}_8) & 
\begin{tikzpicture}
\draw (0,0) node {\scriptsize $\mathbb{F}_{10}$};
\draw (1,0) node {\scriptsize $\mathbb{F}_8$};
\draw (2,0)  node {\scriptsize $\mathbb{F}_6$};
\draw (3,0) node {\scriptsize $\mathbb{F}_4$};
\draw (4,0)  node {\scriptsize $\mathbb{F}_2$};
\draw (5,0) node {\scriptsize $\mathbb{F}_0$};
\draw (6,0)  node {\scriptsize $\mathbb{F}_2$};
\draw (7,0)  node {\scriptsize $\mathbb{F}_2$};
\draw (5,1) node {\scriptsize $\mathbb{F}_2$};
\draw  (0.3, 0) -- (0.7,0); 
\draw  (1.3, 0) -- (1.7,0); 
\draw  (2.3, 0) -- (2.7,0); 
\draw  (3.3, 0) -- (3.7,0); 
\draw  (4.3, 0) -- (4.7,0); 
\draw  (5.3, 0) -- (5.7,0); 
\draw  (6.3, 0) -- (6.7,0); 
\draw  (5, 0.3) -- (5,0.7); 
\end{tikzpicture}
& \mathbb{Z}_{12} \cr\hline
\end{array}
$$
\caption{NHCs, surface configurations and the 1-form symmetry computed from the intersection matrix. 
\label{tab:NHCS}}
\end{table}

There are also NHCs with reducible gauge group on a collection of curves in the base:
\be\label{MultiNHC}
\begin{tabular}{c||c|c|c}
$-n$ & $(-3)(-2)$& $(-3)(-2)(-2)$& $(-2)(-3)(-2)$ \cr \hline
Gauge algebra $\mathfrak{g}$ & $\mathfrak{g}_2\oplus \mathfrak{su}(2)$ & $\mathfrak{g}_2\oplus \mathfrak{sp}(1) \oplus \emptyset$  
&$\mathfrak{su}(2)\oplus \mathfrak{so}(7) \oplus\mathfrak{su}(2)$  \cr \hline
Matter representation & $(\bm{7} \oplus \bm{1} , {1\over 2} \bm{2})$ & 
$(\bm{7} \oplus \bm{1} , {1\over 2} \bm{2})$   &  $ (\bm{1}, \bm{8},  {1\over 2} \bm{2})\oplus (  {1\over 2} \bm{2} ,\bm{8}, \bm{1})$
  \cr \hline 
Ab$[\mathcal{C}]$ & $\mathbb{Z}_5$ &  $\mathbb{Z}_7$ &  $\mathbb{Z}_8$ \cr \hline 
$\Gamma^{\text{gauge}}$ & $1$& $1$& $\mathbb{Z}_2$ \cr 
\end{tabular}
\ee
The last line is the 1-form symmetry that the gauge theory with matter preserves. The 1-form symmetry of the theory in 5d is then 
\be
\Gamma = \text{Ab}[\mathcal{C}] \oplus \Gamma^{\text{gauge}} \,.
\ee
The order of the defect group is computed as ord$(\mathcal{C})=p$, with $\mathcal{C} = \mathbb{Z}_p$ determined by
\be
{p\over q} = n_1 - {1\over n_2 - {1\over n_3- \cdots}} \,.
\ee
The geometry of the NHCs with multiple curves is discussed in appendix \ref{app:NHCs}, where we confirm the above 1-form symmetry groups, by computing the resolution of the Tate model of the elliptic fibration and computing the intersection matrix between compact divisors and compact curves. 

Whenever we tune the gauge group on a single curve of self-intersection $-n$ to a larger group than the one forced in the NHC, 6d anomaly cancellation requires introducing matter, which breaks the 1-form symmetry that the gauge algebra would contribute (see, \eg, table 3 in \cite{Heckman:2018jxk}). This matter can break the 1-form symmetry that arises from the center of the gauge group. 
The only case where the matter is compatible with some remnant 1-form symmetry is 
\be
n= -4 \qquad \text{with }\mathfrak{g}= \mathfrak{so}(N) + (N-8) \bm{V} \,.
\ee
Field theoretically this would result in a theory with 
\be\label{SOon4}
\Gamma = \mathbb{Z}_4 \oplus \mathbb{Z}_2 \,,
\ee
where the $\mathbb{Z}_2$ is the subgroup of the center of $\mathfrak{so}(N)$ that acts trivially on the vector matter. 
This result can be checked from the geometry by noting that the surfaces are \cite{Bhardwaj:2018vuu}
\be
\begin{tikzpicture}
\draw (0,0) node {$\mathbb{F}_{2}$};
\draw  (0.3, 0) -- (1.7,0); 
\node at  (0.35, 0.2){\scriptsize{$e$}};
\node at  (1.65, 0.2){\scriptsize{$h$}};
\draw (2,0) node {$\mathbb{F}_0$};
\draw  (2.3, 0) -- (3.7,0); 
\node at  (2.35, 0.2) {\scriptsize{$h$}};
\node at  (3.65, 0.2) {\scriptsize{$e$}};
\draw (4,0)  node { $\mathbb{F}_2$};
\draw  (4.3, 0) -- (4.6,0); 
\node at (4.4, 0.2) {\scriptsize{$h$}};
\draw (4.9,0) node {\scriptsize $\cdots $};
\draw  (5.05, 0) -- (5.5,0); 
\node at (5.4, 0.2) {\scriptsize{$e$}};
\draw (6,0)  node { $\mathbb{F}_{2r-8}$};
\draw (6.5,0) -- (7.6,0);
\node at  (6.6, 0.2){\scriptsize{$h$}};
\node at  (7.5, 0.2){\scriptsize{$e$}};
\draw (8.3,0) node { $\mathbb{F}^{4r -16}_{2r-6}$};
\draw (8.0,2)  node { $\mathbb{F}_{2r-6}$};
\draw (2,2)  node {$\mathbb{F}_2$};
\draw (6, 0.3)  -- (7.7, 1.7) ;
\node at  (6.15, .65){\scriptsize{$h$}};
\node at  (7.4, 1.65){\scriptsize{$e$}};
\draw  (8, 0.3) -- (8,1.7); 
\node at  (8.2, 1.6){\scriptsize{$f$}};
\node at  (8.8, 0.5){\scriptsize{$f-x_1-x_2$}};
\draw  (2, 0.3) -- (2,1.7); 
\node at  (2.15, 0.5){\scriptsize{$h$}};
\node at  (2.15, 1.7){\scriptsize{$e$}};
\draw  (8.5, 1) node {\boxed{$\scriptsize $_{2r -8}}};
\end{tikzpicture}
\ee 
where the boxed entry $2r-8$ means the surfaces are glued with this multiplicity along curves. Application of the intersections of the (blown up) Hirzebruch surfaces  confirms the 1-form symmetry (\ref{SOon4}).
For all other higher self-intersection number curves $n>2$ with gauge groups $\mathfrak{g} \supset \mathfrak{g}_{\text{NHC}}$ the 1-form symmetry is $\Gamma = \mathbb{Z}_n$. 

An alternative check is to compute the resolution of the elliptic fibration with an $I_p^*$ singularity tuned above $(-n)$, similar to the analysis in appendix \ref{app:NHCs}. The resolution for both codimension one and two (in the base) singularities was obtained in \cite{Lawrie:2012gg}. The starting point is the Tate form \cite{Bershadsky:1996nh, Katz:2011qp} 
\be
y^2 + b_1 x y + b_3 y = x^3 + b_2 x^2 + b_4 x + b_6 \,,
\ee
and with the varnishing orders $\text{ord}_{U=0} (b_i) = (1,1,k, 2k+1)$ for $I_{2k-3}^{*s}$ corresponding to $SO(4k+2)$ (a similar analysis for the other groups). $U=0$ here corresponds to the NHC curve with self-intersection number $-4$. Resolving the singularity results in the Cartan divisors $D_{\alpha_i}$, labeled by the roots of $SO(4k+2)$, which intersect in the Cartan matrix, and correspond to the compact surfaces $S_i$. The curves are obtained as complete intersections from the ambient space, and are given by the rulings of the Cartan divisors, as well as additional curves above the codimension two loci, where the vector matter is located. 
Intersecting the curves and compact surfaces results in the matrix $\mathcal{M}_4$, which confirms the above 1-form symmetry.

\subsection{Non-Minimal Conformal Matter}

A simple class of models are the conformal matter theories in 6d. The minimal conformal matter theories of type $(G_1, G_2)$  have a smooth base $B_2$, and so a trivial defect group, but non-trivial flavor symmetry $G_1 \times G_2$. The 6d 2-form symmetry and 1-form symmetries are trivial and upon reducing to 5d the theories indeed do not have any 1-form symmetry. Only after decoupling hypermultiplets does a 1-form symmetry emerge -- e.g. starting with the $(D_N, D_N)$ minimal conformal matter theories and decoupling, results in pure SYM theories $SU(N-2)_k$, $k=N-1, \cdots, 0$. 

 Non-minimal conformal matter theories are non-very Higgsable, i.e. have a singular base.  
The compactification of a NHV theory on $S^1$ results in a 5d SCFT coupled to a 5d $\mathcal{N}=1$ vector multiplet, where the type of the vector multiplet is determined from the tensor branch geometry of the 6d theory \cite{Ohmori:2015pia}. 
For $(G,G)$ non-minimal conformal matter, the flavor symmetry is $G\times G$ and a tensor branch geometry is given by 
\be
[G]-\stackrel{\mathfrak{g}}{2}- \cdots- \stackrel{\mathfrak{g}}{2}-[G]\,.
\ee
There are $N-1$ $(-2)$ curves in the base, which is a base  of generalized type $A$. Dimensional reduction on $S^1$ results in a 5d SCFT, coupled to an $SU(N)$ vector multiplet \cite{Ohmori:2015pia, Apruzzi:2019kgb}. 
The SCFT part (and its IR description as a quiver) does not carry a higher-form symmetry, as the theory has hypermultiplets in the fundamental. The IR descriptions are quivers in terms of the affine Dynkin diagrams of $G$ with $SU(N d_i)$ nodes, where $d_i$ are the Dynkin labels of $G$, with $N$ fundamentals attached at at least one $SU(N)$ node (see, \eg,  table 3 of \cite{Apruzzi:2019kgb}):
\be
\ba
(A_{n-1}, A_{n-1}):& \qquad [N] - \bm{A_{n-1}} - [N ]\cr 
(D_n, D_n): & \qquad [N] - \bm{D_n}^N \cr 
(E_n, E_n): & \qquad [N] - \bm{E_n}^N \,,
\ea
\ee
where $\bm{G}^N$ corresponds to the quiver that is the Dynkin diagram of $G$ with each node of type $SU(N d_i)$. 
The vector multiplet corresponds to a pure $G$ SYM theory, which has ${Z}_G$ 1-form symmetry, which correctly reproduces the contribution of the 6d defect group to the 5d 1-form symmetry, as in \eqref{5d6d1form}. 

\section{General M-theory Compactifications}
\label{sec:gen}

In section \ref{sec:higherformmtheory} we focused on the case of compactifications of M-theory on Calabi-Yau threefolds to obtain 5d $\cN=1$ models.  However, the main argument there, which used the effective abelian description corresponding to a phase of the theory arising from a smooth compactification manifold, holds with straightforward modifications on more general compactification manifolds. 
To have control of the effective theory it is useful to keep the compactifications supersymmetric. So the cases we have in mind here are M-theory on Calabi-Yau four- and five-folds to 3d $\mathcal{N}=1$ and 1d $\mathcal{N}=2$, as well as exceptional holonomy manifolds $G_2$ and Spin$(7)$ to 4d $\mathcal{N}=1$ and 3d $\mathcal{N}=1$ respectively.

We expect that, for smooth compactification manifolds, the low energy description is given by an abelian gauge theory, with $U(1)$ gauge fields associated to the 2nd cohomology generators, and with matter content given by wrapping M2 branes on 2-cycles.  Since these were the only ingredients we used to derive the 1-form symmetry in the Calabi-Yau threefold case, we may make essentially the  same argument for more general compactification  manifolds.  After appropriately modifying the degrees of the various homology groups in the arguments leading to \eqref{gammadef} and \eqref{gammadef2}, we find 
\be \label{gammadefgen} 
\Gamma = H^{\partial}_2(M_d)/ \text{im} f_2  \; = \;  H^{d-2}(M_d)/\text{im}(\hat{f}_{d-2}) \;. 
\ee
This can also be computed in terms of  the intersection matrix of $2$-cycles and $d-2$ cycles, as described in section \ref{sec:higherformmtheoryint}.  Note that in general compactifications there may not be supersymmetric 2-cycles. \Eg, in $G_2$ holonomy, the only calibrated cycles are 3- and 4-cycles, and the Wilson loops would not be supersymmetric. M-theory on $G_2$-manifolds results in 4d $\mathcal{N}=1$ supersymmetric theories, which indeed do not have supersymmetric Wilson loops.

To give further evidence for this formula, we now turn to  an alternative argument, which considers the fluxes of the M-theory $C$-field at the asymptotic boundary of spacetime.  A similar approach was used in \cite{Garcia-Etxebarria:2019cnb} to understand 2 form symmetries in F-theory compactifications, and we comment on the relation to their results.

\subsection{Asymptotic Fluxes in M-theory Compactifications}

An alternative perspective on the higher-form symmetry in M-theory compactifications of general dimensions can be obtained by studying electric and magnetic fluxes for the 3-form $C$-field of M-theory.  In particular, the choice of higher-form symmetry can be traced to a non-canonical choice of boundary condition for torsion fluxes of this field.  A similar perspective was taken in \cite{Garcia-Etxebarria:2019cnb} in the case of type IIB string theory.\footnote{See also \cite{Gukov:1999ya}, for a related discussion in the context of Calabi-Yau 4-folds.}

When placing M-theory on a non-compact 11d spacetime, $X$, we must also specify the asymptotic behavior of the $C$-field.  Specifically, there are various superselection sectors of the theory which can be labeled by the asymptotic ``electric'' and ``magnetic'' fluxes of the $C$-field, which, to a first approximation, take values in $H^7(\partial X)$ and $H^4(\partial X)$, respectively.  However, there are several caveats to this statement.

First, we expect that the correct mathematical formulation of the $C$-field fluxes would not be in terms of ordinary cohomology classes.  In type II string theory, it is known that the proper formulation of RR fluxes is in terms of the K-theory, rather than cohomology, of spacetime \cite{Moore:1999gb,Freed:2006yc}.  For the purposes of classifying the boundary fluxes, however, it was argued in \cite{Garcia-Etxebarria:2019cnb} that in many examples the K-theoretic analysis agrees with that of ordinary cohomology, and so it is sufficient to consider the latter.  Since we do not know of  the  proper mathematical formulation of  the $C$-field in M-theory, we will instead consider the fluxes as valued in ordinary cohomology classes, and we leave it as an open question to find the appropriate modifications to the discussion below using a more precise formulation. 

Second, recall from the exact sequence in \eqref{h4les} there is a map,
\be \label{asflux} \hat{g}_4: H^4( X) \rightarrow H^4(\partial X) \;. \ee
Only elements in the image of this map can be created by excitations in the bulk, so the only possible fluxes at the boundary lie in the subgroup (in the notation of \eqref{h4les}),
\be \label{asfluxsg}\text{im} \; \hat{g}_4 \cong \text{ker} \; \hat{h}_5 \subset H^4(\partial X)  \;, \ee
and similarly for $H^7(X)$.

Finally, as argued in \cite{Freed:2006ya}, in the case where $\partial X$ has torsion in its 4th cohomology group (and so, by Poincar\'e duality, also in its 7th cohomology group), it is not possible to simultaneously specify the torsion part of both $H^4(\partial X)$ and $H^7(\partial X)$.  A similar issue arose in \cite{Garcia-Etxebarria:2019cnb} in type IIB string theory, and these issues are related via the circle compactification considered in section \ref{sec:6dto5d}.

Let us now specialize to a spacetime of the form
\be X = S_D \times M_d , \;\;\;\; D+d=11 \;. \ee
As mentioned above, we may take the following ansatz for the $C$-field,
\be C  = \sum_a A_a \wedge \omega_ a , \ee
where $\omega_a$ runs over a basis of the 2nd cohomology of  $M_d$, but now we do not demand that $\omega_a$ have compact support.   Suppose we take $A_a$ to lie in a bundle labeled by $c_a \in H^2(S_D)$.  Then the $C$-field has a flux given by
\be \sum_a c_a  \otimes  \omega_a \in H^2(S_D) \otimes H^2(M_d) \subset H^4(X)  \;.\ee
This then projects to a class in $H^4(\partial X)$, as in \eqref{asflux}, defining the  asymptotic magnetic flux of the $C$ field.  Note this indeed lives in the subgroup \eqref{asfluxsg}.  In fact, since $S_D$ is compact, $\hat{g}_4$ only acts on the component from $M_d$, and we may use
\be \hat{g}_2(H^2(M_d))  \cong \Gamma \oplus \Z^f \subset H^2(\partial M_d) \;, \ee
where we have used the Poincar\'e-Lefschetz dual of \eqref{h4gammadef}.  Then we may label the fluxes by
\be \label{asfluxdecomp} H^2(S_D) \otimes (\Gamma \oplus \Z^f) \cong H^2(S_D)^f  \oplus H^2(S_D,\Gamma)  \;. \ee

Let us consider the two factors in this direct sum in turn.  For the first factor, these correspond to 2-cocycles, $\omega_a$, which have non-compact support.  The corresponding gauge fields can be thought of as background, rather than dynamical, gauge fields, and couple to flavor symmetries of the effective theory on $S_D$.  Hence we identify $f$ with the rank of the flavor group, as claimed in the previous subsection, with $H^2(S_D)^f $  labeling the GNO fluxes of the background flavor gauge fields.

On the other hand, the second factor in \eqref{asfluxdecomp} corresponds to cocycles, $\omega_a$, which have non-compact support, but such that some power of them have (some gauge-equivalent representative with) compact support.  For example, suppose we take a $U(1)$  gauge field, $A_a$, associated to such a  cocycle, $\omega_a$ and suppose  that $n \omega_a$ is gauge equivalent to a compact cocycle.  If $S_D \cong \R^D$, then $A_a$ is topologically trivial, and so  we may write
\be A_a \otimes \omega_a  = (\frac{1}{n} A_a) \otimes (n \omega_a) \;, \ee
which is equivalent to a gauge field with compact support, and so does not influence  the asymptotic behavior of the $C$ field.  On the other hand, if $S_D$ is topologically non-trivial, then there may be an obstruction to defining $\frac{1}{n} A_a$ as living in a well-defined gauge bundle.  This obstruction lives in
\be H^2(S_D,\Z_n) \;. \ee
More generally, we see that the obstruction to redefining a gauge field as a compact gauge field lives in the group
\be \label{h2gamm} H^2(S_D,\Gamma) \;,\ee
which is precisely the first factor in \eqref{asfluxdecomp}.  In other words, a non-zero choice in this group instructs us not to integrate over ordinary $U(1)^r$ gauge fields on $S_D$, but rather to those with a given obstruction in \eqref{h2gamm}.  This is precisely the prescription for turning on a background $2$-form gauge field coupled  the electric 1-form symmetry group $\Gamma$ of the gauge theory.   Thus we learn that the factor $H^2(S_D,\Gamma)$ in \eqref{asfluxdecomp} labels the 1-form symmetry backgrounds.

Finally, we observe that we have the option to gauge this 1-form symmetry, as discussed in section \ref{sec:hfsgen}.  This simply means summing over the possible values of the background field, \ie, the elements in $H^2(S_D,\Gamma)$.  The resulting object is then labeled by the Pontryagin dual group, $H^{D-2}(S_D,\widehat{\Gamma})$, where $\widehat{\Gamma}\equiv \text{Hom}(\Gamma,U(1))$.  But note that, thanks to the non-degenerate linking pairing,
\be \ell: TH^2(\partial M_d) \times TH^{d-2}(\partial  M_d) \rightarrow U(1) \;, \ee
where $TH^i$ denotes the torsion subgroup, we may identify
\be H^{d-2}(M_d) \cong \widehat{\Gamma} \;. \ee
Then we  claim that, after gauging, the backgrounds are labeled by fluxes in
\be H^{D-2}(S_D,\widehat{\Gamma}) \cong H^{d-2}(\partial M_d) \otimes TH^{d-2}(\partial M_d) \subset TH^7(\partial X)  \;. \ee
Thus the choice of whether to gauge the 1-form symmetry or not corresponds to the choice of whether to refine by the torsion part of $H^4(\partial X)$ or  $H^7(\partial X)$.  Due to  the non-commutativity of these fluxes \cite{Freed:2006ya}, we may not have both fluxes at once, but must take one choice or  the other (or intermediate choices), and these correspond to the various choices of gauging subgroups of the 1-form symmetry.

\subsection{Surface Operators}

The above argument shows that the labels of an observable implied by the properties of the asymptotic flux agree with those we expect from the higher-form symmetries derived in the previous subsection.  However, a more direct way to see the higher-form symmetries is to construct the topological surface operators measuring  the charge under these symmetries.  Let us briefly describe this construction.

In M-theory, associated to the  3-form gauge field, $C$, we may define the following ``electric'' and ``magnetic'' charge operators,
\be \label{ueumdef} U_E[\alpha \Omega_7] = e^{2 \pi i \alpha \int_{\Omega_7} G_7}, \;\;\;\;\;\; U_M[\beta \Omega_4] = e^{2 \pi i \beta \int_{\Omega_4} G_4} \;, \ee
where $\Omega_4$ and $\Omega_7$ are 4- and 7-cycles in spacetime, and:
\be G_4 = dC, \;\;\; G_7 = \star d C \ee
Since there are both dynamical electrically and magnetically charged objects of unit charge (M2 branes and M5 branes, respectively), $\alpha$ and $\beta$ above must be integers for these to be well-defined.  Then for compact cycles, these operators are actually trivial.  However, the situation is more subtle for non-compact cycles, which we take to mean relative cycles in $H^\partial_i(X)$.

For concreteness, let us consider a spacetime of the form $\R^D \times M_d$, where $M_d$ is the compactification manifold, which we take to be non-compact, and $d+D=11$.  We consider wrapping an M2 brane and a $U_E$ operator on curves,
\be 
\text{M2} \;\; \rightarrow \;\; \sigma_1 \times \omega_2 , \;\;\;\;\; U_E \;\; \rightarrow \;\; \sigma_{D-2} \times \omega_{d-2} \,,
\ee
where $\sigma_1$, $\sigma_{D-2}$ are compact surfaces in $\R^D$ of dimension 1 and $D-2$, respectively, and $\omega_2,\omega_{d-2}$ are non-compact cycles in $M_d$, representing certain relative homology classes.  In the effective $D$-dimensional theory, we interpret this setup as a 1d line operator supported on $\sigma_1$, along with a topological surface operator supported on $\sigma_{D-2}$.  We claim the latter is precisely the charge operator for the 1-form symmetry of this theory discussed above.  This implies it should have non-trivial correlation function with the line operator depending on its charge and the linking number of $\sigma_1$ and $\sigma_{D-2}$.

To see this, first note that the M2 brane creates a unit $G_7$ flux, so if we integrate this flux over a cycle which is linking the M2 brane with linking number $\ell$, we find the flux is equal to $\ell$.  Normally this linking number is an integer, and then, since $\alpha$ is an integer in \eqref{ueumdef}, the expectation value will be trivial.  However, in the present case, the linking number can be computed as
\be 
 \ell( \sigma_1 \times \omega_2 , \sigma_{D-2} \times \omega_{d-2} ) = \ell(\sigma_1,\sigma_{D-2}) ( \omega_2 \cap \omega_{d-2} )\,, \ee
where the linking number on the RHS is computed in $\R^D$, and the intersection number in $M_d$.  The former is an integer, but the intersection number of two non-compact cycles is more subtle to define.  We claim that a natural definition assigns it not an integer, but a rational number (see, \eg,  appendix B of \cite{Closset:2018bjz}).  Specifically, the fractional part of this rational number can be computed using the maps $g_i$ in \eqref{h2les} as
\be \omega_2 \cap \omega_{d-2}  \; = \;  L_1(g_2(\omega_2),g_{d-2}(\omega_d-2))  \; (\text{mod} \; \Z)\,, \ee
where
\be L_i: \quad TH_i(\partial M_d) \times TH_{d-i-2}(\partial M_d) \ \rightarrow \ \bbQ/\Z 
\ee
is the linking form on the boundary, $\partial M_d$.  But this is precisely the pairing which determines the charge of the line operator under the higher-form symmetry.  Thus the expectation value the operators in the setup above contains a factor
\be e^{2 \pi i  \ell(\sigma_1,\sigma_{D-2}) L_1(g_2(\omega_2),g_{d-2}(\omega_d-2))} \ee
which is the expected behavior for the correlation function of a 1-form charge operator with a line operator.  

Finally, if we place the surface operators on topologically non-trivial cycles in a compact spacetime manifold, $S_D$, we expect this corresponds to turning on a background 2-form gauge fields coupled to the 1-form symmetry.  To argue for this, note that, \eg, the $U_E$ operator inserts a delta function $7-$form flux transverse to itself, and so one finds that for a non-compact charge operator
\be 
\ba
U_E[\Omega_7] \; \text{operator insertion}  \;\;\; &\leftrightarrow \;\;\;   \text{asymptotic discrete flux $\mu_7(\Omega_7) \in H^4(\partial X)$ for $G_4$}  \cr
\label{gfu}  U_M[\Omega_4] \; \text{operator insertion} \;\;\; &\leftrightarrow \;\;\;  \text{asymptotic discrete flux $\mu_4(\Omega_4) \in H^7(\partial X)$ for $G_7$}  \,,
\ea
\ee
where we defined a map, $\mu_i$,
\be \mu_i : H^\partial_i(X) \rightarrow H^{11-i}(\partial X)  \;, \ee
by first applying the map $g_i$ in \eqref{h2les}, to obtain an element of $H_{i-1}(\partial X)$, and then applying Poincar\'e duality on $\partial X$.   If we now wrap $\Omega_7$ on a $D-2$ cycle in $S_D$, this implies it creates  the expected asymptotic flux corresponding to a background 2-form gauge field, as in the previous subsection.

\subsection{M-theory on $G_2$-Spaces and 4d $\mathcal{N}=1$ Gauge Theories}

As an example, let us consider compactifying M-theory on a 7-dimensional $G_2$ manifold, leading to an effective 4d theory with $\cN=1$ supersymmetry. 

\subsubsection{Pure SYM and Mass Gap}

Starting with the Bryant-Salamon metric \cite{BryantSalamon} on the local $G_2$ {manifold} $\mathbb{R}^4 \times S^3$, by quotienting with a discrete group of ADE type, we are left with a $G_2$ holonomy {singular space}, $X_7$, whose M-theory compactification gives rise to the  4d $\mathcal{N}=1$ SYM theories with gauge group $G= ADE$ \cite{Atiyah:2000zz, Acharya:2000gb}.  This theory has only adjoint matter, and so preserves a 1-form symmetry $\text{Ab}[\Gamma_{ADE}]$, equal to the center of the corresponding simply-connected ADE Lie group.  
This geometry characterizes the UV of the gauge theory, and is given by a  K3-fibration over $S^3$ and is simply-connected. 
We can compute the 1-form symmetry from the geometry using \eqref{gammadefgen} and (\ref{gammah1dm}), and find 
\be \label{GammaX7}
\Gamma = H_1(\partial X_7) =\text{Ab}[ \Gamma_{ADE}]\,,
\ee
where we have used that $X_7$ is simply-connected.
As discussed in \cite{Atiyah:2000zz} this model is related by a geometric transition, a $G_2$-flop, to a theory on the
space $\widehat{X}_7$ which is an $\R^4$ bundle over $S^3/\Gamma_{ADE}$. 
$X_7$ and $\widehat{X}_7$ are connected by varying the complexified coupling, $\tau$. 
The theory exhibits confinement, and the 1-form symmetry is unbroken in that phase.  
The confining strings were described in \cite{Acharya:2001hq} as M2 branes wrapping 1-cycles in the geometry. These are precisely charged under {the} $\Gamma$ {given} in (\ref{GammaX7}). The mass gap can be understood from considering the metric fluctuations of the $G_2$-manifold $\widehat{X}_7$ \cite{Acharya:2004qe}, which are not normalizable and thus the spectrum has a mass gap.

Another simple class of examples of manifolds with (at most) $G_2$ holonomy are simply products of a Calabi-Yau threefold and $S^1$.  These lead to 4d theories with $\cN=2$ supersymmetry.  For simply connected Calabi-Yau manifolds, the formula \eqref{gammadefgen} gives the same 1-form symmetry as the 5d theory, which is as expected as we expect the 5d 1-form symmetry to reduce to a 1-form symmetry in 4d. Unlike for $\cN=1$ compactifications, in this case there is a notion of calibrated 2-cycles on which we may wrap M5 branes, and so correspondingly a notion of supersymmetric loop operators, which can be charged under the 1-form symmetry.

\subsubsection{Models with Chiral Matter}

Extending this to theories with matter, recall that chiral matter in 4d is created by conical singularities of codimension 7 in the $G_2$ -holonomy manifold \cite{Acharya:2001gy, Berglund:2002hw}. The local geometry that we wish to model is $ALE$-fibration over $M_3$, which degenerates at points. 
To realize fundamental matter for $SU(N)$, the geometry corresponds to an unfolding of an $SU(N+1)$ singularity using the hyper-K\"ahler quotient construction.  The geometry constructed in \cite{Acharya:2001gy} has $M_3= \mathbb{R}^3$, and shows that chiral matter in the $\bm{N}_{+}$ of $U(N)={SU(N)\times U(1)\over \mathbb{Z}_N}$
is realized by the cone over a weighted projective space
\be
X_7 = C \left(\mathbb{CP}^{1,1,N, N,}\right) \,,
\ee
It follows that $\Gamma= H_1 (\partial X_7, \mathbb{Z}) = H_1 (\mathbb{CP}^{1,1,N, N,}) = 0$ (e.g. \cite{Kawasaki}). 
Similarly, the unfolding of $SU(N+1) \rightarrow SU(p) \times SU(q) \times U(1)$ is captured by the cone over $\mathbb{CP}^{p,p,q,q}$, which also has $\Gamma = 0$ -- consistently with the field theory expectation of bi-fundamental matter for $SU(n)$ groups.

Generalizing to $\text{Spin}(2n)$ gauge theories, the field theory expectation is that there is a $\Z_2$ 1-form symmetry preserved for vector matter. In \cite{Berglund:2002hw} the corresponding hyper-K\"ahler quotient was discussed and shown to be a $\mathbb{Z}_2$-quotient of the one for $A_n$
\be
x y = z^{2n}\,, \qquad (x,y,z) \rightarrow (y,x, -z) \,.
\ee
The Higgsing
\be
\text{Spin}(2n+2) \ \rightarrow\  {\text{Spin}(2n) \times \text{Spin}(2)\over \mathbb{Z}_2} \,,
\ee
describes the 4d theory ${\text{Spin}(2n) \times \text{Spin}(2)\over \mathbb{Z}_2}$  and a vector which is charged $+2$ under the $\text{Spin}(2)$. This theory preserves a $\mathbb{Z}_2$ 1-form symmetry, which is confirmed from the geometry, as the $\mathbb{Z}_2$ quotient the space has now non-trivial first homology and $\Gamma= \mathbb{Z}_2$. Similar considerations can be studied using the Higgs bundle approach to $G_2$-holonomy manifolds  \cite{Pantev:2009de, Braun:2018vhk}, to generalize this to theories with more general matter content.

\section{Summary and Outlook}
\label{sec:Conclusions}

In this paper we studied 1-form symmetries in 5d gauge theories and identified the geometric counterpart in their M-theory realization. 
Whenever possible we cross-checked the 1-form symmetry in all available descriptions: gauge theory, M-theory geometry, and directly in terms of the geometry of surfaces that realize the gauge theory in M-theory -- and find agreement. 
1-form symmetries have been observed to increase (and never decrease) under RG-flows. In particular starting with a 6d model, the circle compactification to 5d (and mass deformation) yields a theory with 1-form symmetry at least {as large as} that in 6d, but generically enhanced by the 2-form symmetry in 6d, which is the defect group in 6d. Note that not all dimensional reductions correspond to 5d SCFTs, \eg, when the theory has a non-trivial defect group in 6d (and is a so-called non-very Higgsable theory in 6d), the theory in 5d is an SCFT coupled to a gauge sector \cite{Ohmori:2015pia, Apruzzi:2019kgb}. The 1-form symmetry of the combined theory agrees with the one in 6d. The 5d SCFT can thus in general have a smaller 1-form symmetry than the 6d theory.

There are various extensions {which might be made} of this work. 
The most interesting direction is to construct observables, which are sensitive to the 1-form symmetry -- such as was discussed in 3d in \cite{Eckhard:2019jgg} in the context of the Witten index.  One class of such observables would be supersymmetric partition functions on spaces with non-trivial $H^2(S_{(5)})$; \eg, Sasaki-Einstein manifolds (see  \cite{Qiu:2016dyj} for a review), $\Sigma_{g_1} \times \Sigma_{g_2} \times S^1$ \cite{Hosseini:2018uzp}  or $S^3_b \times \Sigma_g$ \cite{Crichigno:2018adf}, should be of interest. 

It would also be interesting to define the 1-form symmetry more intrinsically and develop a method to determine how it changes along mass deformations/RG-flows. For the 0-form symmetries, \ie, flavor symmetries, this is achieved by the combined fiber diagrams (CFDs) \cite{Apruzzi:2019vpe, Apruzzi:2019opn, Apruzzi:2019enx, Apruzzi:2019kgb, Eckhard:2020jyr} {which} encode the enhanced flavor symmetry of the UV fixed point (as well as all possible IR descriptions) by embedding the associated flavor symmetries. Furthermore, {those diagrams}
encode all mass deformations, which allows tracking of the flavor symmetry along RG-flows. Much like the 1-form symmetry, the 0-form symmetry and CFDs do not depend on internal flops, \ie, choosing different Coulomb branch chambers for the same IR gauge theory description. It would be desirable to construct a similar structure that encodes the 1-form symmetry and its changes under RG-flows, and furthermore to combine this structure with the CFDs. Note that the CFDs are non-trivial whenever there are flavors, and become more trivial towards the bottom of the RG-flow tree. In contrast, the 1-form symmetry becomes non-trivial once the flavors are decoupled.

For some theories we do not have a gauge theoretic description, such as the toric $B_N$ theories and the $\mathbb{P}^2$ theory at rank 1, which do not have any weakly coupled gauge theory description, although on the Coulomb branch we have some support for the 1-form symmetries in terms of that of the abelian gauge theory. 
It would be interesting to test these by further compactification to lower dimensions, \eg, 4d or 3d, 
where the theories may have a Lagrangian description and a field theory analysis can confirm the 1-form symmetry. 

Finally, we have only briefly discussed the generalizations, and a more in depth analysis for higher dimensional Calabi-Yau manifolds, {as well} as $G_2$ or Spin$(7)$ manifolds, would be of interest. Furthermore we assumed that {our} Calabi-Yau {threefolds} have no torsion. It would be interesting to extend the analysis by relaxing this condition.

Another generalization is to consider non-supersymmetric theories, such as those proposed in  \cite{BenettiGenolini:2019zth}, where the rank 1 $SU(2)_0$ theory 
is mass deformed into a non-supersymmetric theory, which then has a conjectural UV completion in a 5d non-supersymmetric CFT. The mass deformations do not affect the theta angle, which remains $0$, and thus we expect the 1-form symmetry to be present also in the non-supersymmetric case. This would be interesting to extend and consider in other instances, \eg, for $SU(N)_{kN}$ pure SYM theories in 5d, which for certain $k$ (\eg, $SU(3)_3$) also exhibit a global $SU(2)_F$ 0-form symmetry.

\subsection*{Acknowledgements}

We thank Fabio Apruzzi, Lakshya Bhardwaj, Sebastjan Cizel, Cyril Closset, Julius Eckhard, Ken Intriligator, Shlomo Razamat, Nathan Seiberg, James Sparks, Washington Taylor, Yi-Nan Wang  for discussions. SSN thanks the KITP for hospitality.
DRM and BW are supported by the Simons Foundation Grant \#488629 (Morrison) and the Simons Collaboration
on Special Holonomy in Geometry, Analysis and Physics.
The work of SSN is supported by the ERC Consolidator Grant number 682608 ``Higgs bundles: Supersymmetric Gauge Theories and Geometry (HIGGSBNDL)''.

\appendix

\section{1-Form Symmetries from Intersecting Surfaces} 
\label{app:1formsurface}

\subsection{Gauge theories from intersections}

As described, \eg, in \cite{Bhardwaj:2019ngx}, we may engineer any allowed 5d $\cN=1$ gauge theory as a collection of intersecting surfaces, as above.  We emphasize at this level we are agnostic about whether the given gauge theory has a UV completion as an SCFT, considering it for now only as an effective description.  The prescription is as follows:

\begin{enumerate}
	\item Start with the geometry engineering a pure gauge theory with the desired (semi-simple) group $G$, as described in section 2.4 of \cite{Bhardwaj:2019ngx}.  In particular, this has the property that there are $r \equiv \text{rank}(G)$ divisors, and the matrix:
	\be C_{ij} = - f_i \cdot S_j , \;\;\; i,j =1,...,r\ee
	where $f_i$ is the fiber of $S_i$, is the Cartan matrix of $G$.  Note that for $G=SU(N)$ and $Sp(N)$ we must also choose the geometry to include the appropriate CS term/theta angle.
	\item To add matter, we focus on the case where the matter representation has Dynkin labels at most $1$.  Recall we can identify the Dynkin indices with the simple roots of $G$, which in turn can be identified with the divisors in the pure gauge theory geometry above.  Then for each non-zero Dynkin label, we blow up a curve in the corresponding divisor.  If there is more than one such divisor, we identify these curves by gluing.  We repeat this for all matter representations.
	\item This places us in the phase where the matter has been ``minimally integrated in,'' \ie, the first phase after the one where the mass of the hypermultiplet can be taken to $-\infty$.  Other phases can be obtained by flopping curves, as described below.
\end{enumerate}

Let us now argue that the $1$-form symmetry from the geometry engineering a gauge theory agrees with that read off from the gauge theory description, as described in section \ref{sec:5dhfs} above.  First we must check that the pure gauge theories have the appropriate 1-form symmetry.  For example, consider the geometry engineering $SU(N)_k$ for $N \geq 3$:
\be \mathbb{F}_{N-2-k} \; \;_e\qline_h \; \; \mathbb{F}_{N-4-k} \;  \;_e\qline \cdots  \qline_h \; \mathbb{F}_{4-N-k}  \; \;_e\qline_h \; \; \mathbb{F}_{2-N-k}  \ee
where $h\equiv e+nf$ in $\mathbb{F}_n$.  In general, the curves $e_a$ and $f_a$, as $a=1,,...,r$ ranges over the divisors, generate the curves in the geometry, but, for simple $G$, the gluing curves allow us to eliminate all but one of the $e_a$, say\footnote{For semi-simple $G$, there will be an equivalence class of $e$ curves for each simple factor.}  
$e_1$.
Dropping these redundant rows, the matrix $\Omega$ in this case can be written as:
\be	\Omega =\begin{blockarray}{cccc}
	& \cdots & S_a & \cdots  \\
	\begin{block}{c(ccc)}
		\vdots & & \vdots  & \\
		f_b & \cdots & C_{ab} & \cdots  \\
		\vdots & & \vdots  &  \\
		e_1 & \cdots & \gamma_a & \cdots  \\
	\end{block}
\end{blockarray}  \;. \ee
where  $C_{ab}$ is the Cartan matrix, and we defined $\gamma_a = e_1 \cdot S_a$.  For the example of $SU(N)_k$ above, this is given by:
\be \gamma_a = \begin{cases} N-k-4 & a=1 \\ -N+k+2 & a=2 \\ 0 & \text{else} \end{cases} \;. \ee
Now the Cartan matrix, $C$, of $SU(N)$ can be diagonalized as:
\be D= S  C  T  \ee
where,
\be S = \begin{pmatrix} 
	1 & 1 & 1 & \cdots & 1 \\
	1 & 2 & 2 & \cdots & 2 \\
	1 & 2 & 3 & \cdots & 3 \\
	\vdots & & & & \vdots \\
	1 & 2 & 3 & \cdots & N-1 \end{pmatrix}, \;\;\; D = \begin{pmatrix} 
	1 & 0 & \cdots & 0 \\
	0 & 1 & \cdots & 0 \\
	\vdots &  & & \vdots \\
	0 & 0 & \cdots & N \end{pmatrix},  \;\;\; T = \begin{pmatrix} 
	1 & 0 & 0 & \cdots & -1 \\
	0 & 1 & 0 & \cdots & -2 \\
	0 & 0 & 1 & \cdots & -3 \\
	\vdots & & & & \vdots \\
	0 & 0 & 0 & \cdots & -(N-2) \\
	0 & 0 & 0 & \cdots & 1 \end{pmatrix}  \;. \ee
Recall $\Omega$ has the form,
\be \label{CAform}\Omega = \begin{pmatrix} C \\ A\end{pmatrix} \;, \ee
where  $A$ is an $\ell \times (N-1)$ matrix, in this case with $\ell=1$.  Then if we act on the left with $\begin{pmatrix} S & 0 \\ 0 & 1 \end{pmatrix}$ and on the right with $T$, we obtain
\be \Omega' = \begin{pmatrix} D \\ A T \end{pmatrix} \;, \ee
Since all but the last diagonal entry of $D$ is $1$, we may act again on the left to eliminate all but the last column of $AT$ to get (writing $D$ explicitly),
\be \Omega'' = \begin{pmatrix} 1 & \cdots & 0 \\ \vdots &  & \vdots \\ 0 & \cdots & N  \\ 0 & \cdots  & \alpha \end{pmatrix} \;, \ee
where $\alpha$ is a length $\ell$ column vector with components
\be \alpha_a  =  - \sum_{b=1}^{N-1} b A_{ab}  \; (\text{mod} \; N)  \;, \ee
which may be considered mod $N$, since we can shift by a multiple of $N$ by adding the last row of the diagonalized Cartan matrix.  In the present case, $A_{1,b}=\gamma_b$, and this gives:
\be \Omega'' = \begin{pmatrix} 1 & \cdots & 0 \\ \vdots &  & \vdots \\ 0 & \cdots & N  \\ 0 & \cdots  & -k \end{pmatrix} \;, \ee
Finally, using additional row operations on the last column, we obtain:
\be \Omega''' = \begin{pmatrix} 1 & \cdots & 0 \\ \vdots &  & \vdots \\ 0 & \cdots & \text{gcd}(N,k)  \\ 0 & \cdots  & 0 \end{pmatrix} \;, \ee
which gives, using \eqref{gzfgo},
\be \Gamma \oplus \Z^f = \Z_{\text{gcd}(N,k)} \oplus \Z \;,\ee
as expected.  A similar analysis holds for other pure gauge theories, showing that their 1-form symmetry is equal to their center.\footnote{In the case of $Sp(N)$, the 1-form symmetry is equal to the $\Z_2$ center for $\theta=0$, and is trivial for $\theta=\pi$.}

Next we add matter.  We expect the matter will break the 1-form symmetry of the pure gauge theory to the subgroup of the center acting trivially on the matter.  Continuing with the $SU(N)_k$ example for concreteness, suppose we add matter in a representation with Dynkin labels $\delta_a$, $a=1,...,N-1$, where we take $\delta_a \in \{0,1\}$ as mentioned above.  This is included by blowing up a curve in the corresponding divisors, and adds a row to the matrix $\Omega$ as:
\be	\Omega =\begin{blockarray}{cccc}
	& \cdots & S_a & \cdots  \\
	\begin{block}{c(ccc)}
		\vdots & & \vdots  & \\
		x & \cdots & -\delta_a & \cdots   \\
	\end{block}
\end{blockarray}  \;. \ee
This is again a matrix of the form \eqref{CAform}, and so reasoning as above, we find the Smith normal form can be written as:
\be \Omega'' = \begin{pmatrix} 1 & \cdots & 0 \\ \vdots &  & \vdots \\ 0 & \cdots & N  \\ 0 & \cdots  & \alpha \end{pmatrix} \;, \ee
where $\alpha$ is now a $2$-component column vector, with first entry $-k$, as above, and second entry:
\be \alpha_2  =   \sum_{b=1}^{N-1} b \delta_{b}  \; (\text{mod} \; N) \equiv q_\delta \ee
This is precisely the $\Z_N$ center charge of the representation with these Dynkin labels, and we have correspondingly denoted this by $q_\delta$.  Including $N_f$ total representations, we find that:
\be \Omega'' = \begin{pmatrix} 1 & \cdots & 0 \\ \vdots &  & \vdots \\ 0 & \cdots & N  \\ 
	0 & \cdots  & -k  \\
	0 & \cdots  & q_{\delta_1}  \\
	\vdots &  & \vdots \\
	0 & \cdots  & q_{\delta_{N_f}} 
\end{pmatrix} \;, \ee
and applying further row operations, we finally obtain:
\be \Omega''' = \begin{pmatrix} 1 & \cdots & 0 \\ \vdots &  & \vdots \\ 0 & \cdots & \text{gcd}(N,k,q_{\delta_1},...)  \\ 
	0 & \cdots  & 0 \\
	0 & \cdots  & \vdots 
\end{pmatrix} \;, \ee
so that, from \eqref{gzfgo},
\be \Gamma \oplus \Z^f \cong \Z_{\text{gcd}(N,k,q_{\delta_1},...)} \oplus  \Z^{N_f+1} \;,\ee
which is the expected result.  One may repeat this argument for other gauge groups.

Finally, we should check the 1-form symmetry is the same in other phases of the gauge theory, which are related by flops, and we turn to this next.

\subsection{Invariance under Surface Equivalences and Internal Flops}
\label{app:Inv}

Next we check that the 1-form symmetry is invariant under various equivalences in the geometry.  

The first is (local) $S$-duality, which exchanges the $e$ and $f$ curves in a divisor $\mathbb{F}_0^b$.  In other words, if one of the divisors in our geometry is of this form, we exchange $e$ and $f$ in any gluing curves involving this divisor.  Recall the intersection matrix for curves $(e, f, x_i)$ is
\be \begin{pmatrix} -n & 1 & 0 \\ 1 & 0 & 0 \\ 0 & 0 & -\delta_{ij} \end{pmatrix} \,,
\ee
which is invariant under exchanging $e$ and $f$ for  $n=0$. Thus the matrix $\Omega$ will be covariant under the exchange of $e$ and $f$,\footnote{In other words, the rows of the matrix corresponding to the $e$ and $f$ curves inside this divisor will be swapped, which will not affect the Smith normal form.} and so the final quotient group in \eqref{gzfgo} will be the same.

Next we consider the isomorphism in eqs (2.19)-(2.22) of \cite{Bhardwaj:2019ngx}, which takes $\bbF_n^b \rightarrow \bbF_{n+1}^b$.  This acts on the curves as (here $x_i$ is the curve being transformed, and $x_j$ are the remaining curves)
\be \begin{pmatrix} e \\ f \\ x_i \\ x_j \end{pmatrix}   \rightarrow  \begin{pmatrix} 
	1 & 1 & -1 & 0 \\
	0& 1 & 0 & 0 \\
	0& 1 & -1 & 0 \\
	0& 0 & 0 & 1 \end{pmatrix} \begin{pmatrix} e \\ f \\ x_i \\ x_j \end{pmatrix} \,. \ee
If we denote the matrix above by $A$, then one can verify that after applying this isomorphism, the matrix $\Omega$ transforms as:\footnote{We write this in block form, with the first block acting on the curves in divisor which we apply the isomorphism to, and the second block acting on all the others}
\be \Omega \rightarrow \begin{pmatrix} A^{-1} & 0 \\ 0 & I \end{pmatrix} \Omega \ee
which clearly does not affect the Smith normal form, and so the group in \eqref{gzfgo} is unchanged.

Finally, we consider the effect of a $(-1,-1)$ flop.  Let us first describe how a flop acts in a collection of intersecting surfaces.  We will flop a curve, $\CC$, which may lie in multiple divisors.  We first partition the set of divisors, $\CS$, into three groups, $\CS= A \cup B \cup C$, as follows:

\begin{itemize}
	\item $A$ - those divisors containing the curve, $\CC$, to be flopped.  We assume that (after possibly applying the isomorphisms above) the curve to be flopped in each divisor is one of the blown up curves, $x_i$.  These are then all identified with each other by gluing curves.  Note then that the intersection of each of these divisors with the curve is $-1$.
	\item $B$ - Those divisors, which intersect the curve.  We assume\footnote{There are more general setups, where these can intersect in multiple points, but the argument readily generalizes.} that they all have intersection number $1$.  This implies that $C_{ab} = \cdots - x$, where $a \in A, b \in B$, and $x$ is the curve in $a$ to be flopped.
	\item $C$ - Those divisors, which do not intersect the curve.
\end{itemize}

Then the effect of the flop is essentially to exchange the role of the divisors in sets $A$ and $B$.  Namely, we blow down the curve sitting inside the divisors in $A$, and then blow up a curve in each divisor in $B$, and these blow ups are all identified.  The divisors in $A$ then meet this common curve with intersection number $1$.  The divisors in $C$ are unchanged.  At the level of the gluing curves, we have, schematically:\footnote{Note there may be multiple gluing curves connecting, say, $S_a$ and $S_{a'}$.  The second line says that one of these is gluing $x_a$ to $x_{a'}$, and that gluing curve disappears after the flop.  Similar comments apply to $S_b$ and $S_{b'}$ on the second line.}
\be C_{ab} = C_{ab}^0 - x_a \; \rightarrow  \; C_{ab}^0, \;\;\;\;\;\; C_{ba} = C_{ba}^0 \;  \rightarrow  \; C_{ba}^0 - x_b  \;, \ee
\be C_{aa'} =x_a \;  \rightarrow  0, \;\;\;\;\; C_{bb'} = 0 \; \rightarrow \; x_b \;. \ee

Note that flopping this new curve brings us back to the original geometry.  Note also that if $B$ is empty, the curve $\CC$ is flopped out of the collection of surfaces, but after the flop it still meets the curves in $A$ with intersection number $1$.  This takes us outside the formalism of \cite{Bhardwaj:2019ngx}, as the geometry now involves compact curves not contained in any of the Hirzebruch surfaces.  We may now take a limit where the curve becomes decompactified, and then lose this compact curve from the geometry, taking us to a new geometry which again lies within the formalism of \cite{Bhardwaj:2019ngx}.

In the case described above, where $B$ is empty, the curve flops out of the collection of surfaces, but we still retain a row as in \eqref{omaf}, although this curve does not sit inside any surface.  If we now take a decompactification limit of this curve, we simply lose this row from $\Omega$.  Note that the flop itself does not change the 1-form symmetry, but taking this decompactification limit can do so.
 
\section{NHC details}

\subsection{Quotient space construction}
\label{sec:nhcquotient}

As mentioned in the main text, we may construct the NHC geometries  for even $n$ as a quotient \cite{Heckman:2013pva},
\be X= (\C^2 \times T^2)/\Z_n  \;,\ee
where $\Z_n$ acts as
\be (z_1,z_2,w) \sim (\omega \; z_1, \omega \; z_2, \omega^{-2} \; w)  \;,\ee
where $\omega=e^{2 \pi i/n}$, $z_i$ are the coordinates of $\C^2$, and $w \sim w+1 \sim w+\tau$ is the coordinate on $T^2$.  Here we should choose $\tau$ to be compatible with the action, eg, $\tau=i$ for $n=8$, and $\tau = e^{2 \pi i/3}$ for $n=6,12$.  The boundary of this space is then a quotient:
\be \partial X = (S^3 \times T^2)/\Z_n \,,\ee
which is a $T^2$ fibration over the lens space, $L(n,1)$.

To determine the 1-form symmetry, we must compute the homology of the boundary.  We will focus on computing $H_1(\partial X)$, which is the abelianization of $\pi_1(\partial X)$.  To compute this, we note that the universal cover of $\partial X$ is $S^3 \times \R^2$.  To obtain $\partial X$, we first quotient by $\Z^2$ to obtain $S^3 \times T^2$, and then quotient by $\Z_n$ as above.  Equivalently, we may quotient the universal cover $S^3 \times \R^2$ by a semi-direct product
\be \pi_1(\partial X) = \Z^2 \rtimes_\phi \Z_n \,,\ee
where $\phi:\Z_n \rightarrow \text{Aut}(\Z^2) \cong SL(2,\Z)$ maps the generator of $\Z_n$ to an appropriate element of $SL(2,\Z)$, \ie\footnote{Here we recall it is $\omega^{-2}$ that acts on the torus, so the $SL(2,\Z)$ element has order $n/2$.}
\be n=2 \;\; : \;\;  I = \begin{pmatrix} 1 & 0 \\ 0 & 1 \end{pmatrix} , \;\;\;\;\;
n=4 \;\; : \;\;  \CC = \begin{pmatrix} -1 & 0 \\ 0 & -1 \end{pmatrix} , \;\;\;\;\;
n=6 \;\; : \;\; (\CS \CT)^2 =  \begin{pmatrix} -1 & -1 \\ 1 & 0 \end{pmatrix} , \ee
\be n=8 \;\; : \;\; \CS = \begin{pmatrix} 0 & -1 \\ 1 & 0 \end{pmatrix} , \;\;\;\;\;
n=12 \;\; : \;\; \CS \CT =  \begin{pmatrix} 0   & -1 \\ 1 & 1 \end{pmatrix} . \ee
Then
\be H_1(\partial X,\Z) =  \pi_1(\partial X)^{Ab} = ({\Z^2})_{\Z_n} \oplus \Z_n\,, \ee
where $({\Z^2})_{\Z_n}$ is the quotient of $\Z^2$ by the subgroup generated by elements of the form
\be f(x) - x, \;\;\;\; x \in \Z^2, \;\;\; f \in \phi(\Z_n) \subset \text{Aut}(\Z^2)\,. 
\ee
For example, for $n=4$, this is the image of the matrix
\be \CC - I = \begin{pmatrix} - 2 & 0 \\ 0 & -2 \end{pmatrix}\,, \ee
which implies the quotient $({\Z^2})_{\Z_n}$ is equal to $\Z_2 \oplus \Z_2$.  A similar computation in the other cases gives the following results for $H_1(\partial X,\Z)$:
\be n=2 \;\; : \;\;   \Z_2 \oplus \Z^2 \,,\qquad 
n=4 \;\; : \;\; \Z_4 \oplus \Z_2 \oplus \Z_2, \qquad 
n=6 \;\; : \;\;\Z_6 \oplus \Z_3, \ee
\be n=8 \;\; : \;\; \Z_8 \oplus \Z_2, \qquad 
n=12 \;\; : \;\; \Z_{12} \,.\ee
The 1-form symmetry is then given by the subgroup of this homology group which is trivial when included into the bulk manifold, $X$.  For $n>2$, the bulk manifolds are simply connected, and so this is the entire group.  For $n=2$, the total space is $\C^2/\Z_2 \times T^2$, and so the $\Z^2$ factors above are still non-trivial in the bulk, and only the $\Z_2$ factor remains.  This is indeed the expected result for the 1-form symmetry of these theories. 

\subsection{Multi-Curve NHCs}
\label{app:NHCs}

In this appendix we compute the geometry of surfaces and curves for the multi-curve NHCs in (\ref{MultiNHC}) and determine from the geometry the 1-form symmetry. 

The geometry of the NHC $\mathfrak{g}_2 \oplus \mathfrak{su}(2)$ on $(-3) (-2)$ curves is obtained by resolving the Tate model (a discussion of the Weierstrass model has appeared in \cite{Esole:2018mqb})
\be
y^2 - x^3 -U^3 V^2 w^6 b_6 -U^2 V w^4 x b_4+U^2 V w^3 y b_3-U w^2 x^2 b_2+U w x y b_1 =0 \,.
\ee
Here above $U=0$ corresponds to the curve with self-intersection $-3$, upon which sits the $\mathfrak{g}_2$ with a Kodaira fiber $I_0^{*ns}$, and $V=0$ is the $-2$-curve, with the $\mathfrak{su}(2)$. The base has four coordinates $U, V$ as well as the non-compact divisors $W, Z$, which are linearly arranged as $W, U, V, Z$. Furthermore $c_1(B) = U + V + W + Z$.
The model is readily resolved by
\be
\{x, y, U; u_1\}\,, \ \{ y, u_1; u_2\}\,,\ \{u_1, u_2; u_3 \} \,,\  
\{x, y, V; v_1\} \,,\ 
\ee
The Cartan divisors for the affine $\mathfrak{g}_2$ are $(U, u_3, u_2)$ and for the affine $\mathfrak{su}(2)$ $(V, v_1)$ \,.
These are the compact divisors. The curves are complete intersections in the ambient space and are given by the following pair-wise intersections 
\be\label{NHCCurves}
C_I= \left(\begin{array}{c|c|c|c|c|c|c|c|c|c|c|c|c|c|c|c|c}
 u_2 & u_2 & u_2 & y & V & x & u_3 & u_3 & w & U & U & v_1 & x & y & x & y & v_1 \\
 u_3 & x & v_1 & u_2 & u_2 & u_3 & V & U & U & V & W & V & V & V & v_1 & v_1 & Z \\
\end{array}\right) 
\ee
The matrix of intersections between compact divisors and curves is then 
\be
\mathcal{M}_4 = \left(
\begin{array}{ccccccccccccccccc}
 0 & 0 & 0 & 0 & 0 & 0 & 1 & -1 & -3 & -2 & -2 & 0 & 0 & 0 & 0 & 0 & 0 \\
 -9 & 3 & 0 & 0 & 3 & -3 & -2 & -1 & 0 & 1 & 1 & 0 & 1 & 0 & 0 & 0 & 0 \\
 9 & -6 & 0 & -9 & -6 & 3 & 3 & 0 & 0 & 0 & 0 & 2 & -2 & 1 & 2 & 2 & 0 \\
 3 & -2 & 2 & 1 & -2 & 1 & 0 & 1 & 1 & 0 & 0 & -1 & -1 & -1 & 1 & -1 & 2 \\
 0 & 2 & -2 & 2 & 2 & 0 & 0 & 0 & 0 & 0 & 0 & -3 & 1 & -1 & -5 & -3 & -2 \\
\end{array}
\right)\,,
\ee
which has Smith normal form 
\be
\left(
\begin{array}{ccccccccccccccccc}
 1 & 0 & 0 & 0 & 0 & 0 & 0 & 0 & 0 & 0 & 0 & 0 & 0 & 0 & 0 & 0 & 0 \\
 0 & 1 & 0 & 0 & 0 & 0 & 0 & 0 & 0 & 0 & 0 & 0 & 0 & 0 & 0 & 0 & 0 \\
 0 & 0 & 1 & 0 & 0 & 0 & 0 & 0 & 0 & 0 & 0 & 0 & 0 & 0 & 0 & 0 & 0 \\
 0 & 0 & 0 & 1 & 0 & 0 & 0 & 0 & 0 & 0 & 0 & 0 & 0 & 0 & 0 & 0 & 0 \\
 0 & 0 & 0 & 0 & 5 & 0 & 0 & 0 & 0 & 0 & 0 & 0 & 0 & 0 & 0 & 0 & 0 \\
\end{array}
\right) \,,
\ee
so that the 1-form symmetry is 
\be
\Gamma = \mathbb{Z}_5 \,.
\ee
For the $(-3)(-2)(-2)$ with $\mathfrak{g}_2 \oplus \mathfrak{sp}(1) \oplus \emptyset $ the only difference is that $W=0$ is now a compact curve with self-intersection $-2$, and we get one additional non-compact divisor, say $Y=0$ in the base, with $c_1(B) = U + V+W+Z+Y$. 
In addition to the curves in (\ref{NHCCurves}) we also have 
\be
C_I = (\ref{NHCCurves}) \cup \left(
\begin{array}{c|c|c|c}
 W & W & W & W \\
 v_1 & x & y & V \\
\end{array}
\right) \,.
\ee
The intersection matrix is then 
\be\ba
&\mathcal{M}_4=\cr 
& {\scriptsize
\left(
\begin{array}{ccccccccccccccccccccc}
 0 & 0 & 0 & 0 & 0 & 0 & 1 & -1 & -3 & -2 & 0 & 0 & 0 & 0 & 0 & 0 & 0 & 0 & 0 & 0 & 0 \\
 -9 & 3 & 0 & 0 & 3 & -3 & -2 & -1 & 0 & 1 & 0 & 0 & 1 & 0 & 0 & 0 & 0 & 0 & 0 & 0 & 0 \\
 9 & -6 & 0 & -9 & -6 & 3 & 3 & 0 & 0 & 0 & 0 & 2 & -2 & 1 & 2 & 2 & 0 & 0 & 0 & 0 & 0 \\
 3 & -2 & 2 & 1 & -2 & 1 & 0 & 1 & 1 & 0 & 0 & -1 & -1 & -1 & 1 & -1 & 0 & 2 & 0 & 1 & -2 \\
 0 & 2 & -2 & 2 & 2 & 0 & 0 & 0 & 0 & 0 & 0 & -3 & 1 & -1 & -5 & -3 & 0 & -2 & 2 & 2 & 2 \\
 0 & 0 & 0 & 0 & 0 & 0 & 0 & 0 & 0 & 0 & 0 & 2 & 0 & 1 & 2 & 2 & 0 & 0 & -4 & -6 & 0 \\
\end{array}
\right)}\,,
\ea\ee
whose Smith normal form implies 
\be
\Gamma = \mathbb{Z}_7 \,.
\ee

Finally the Tate model for the (-2)(-3)(-2) with $\mathfrak{su}(2) \oplus \mathfrak{so}(7) \mathfrak{su}(2)$ NHC is 
\be
b_6 \left(-U^2\right) V^4 w^6 W^2-b_4 U V^2 w^4 W x+b_3 U V^2 w^3 W y-b_2 U V w^2 W x^2+b_1 U V w W x y-x^3+y^2=0 \,,
\ee
which is resolved by 
\be
\left\{x,y,U;u_1\right\},\ \left\{x,y,W;w_1\right\},\ \left\{x,y,V;v_1\right\},\ \left\{x,y,v_1;v_2\right\},\ \left\{y,v_1;v_3\right\},\ \left\{v_1,v_3;v_4\right\}\,.
\ee
where the Cartan divisors for the affine algebras $S_i=(V,v_2,v_4,v_3,U,u_1,W,w_1)$.
The set of curves are not only those with $S_i \cdot_Y \zeta$, where $\zeta$ is any of the ambient space sections, but also triple intersections of ambient space sections, that result in curves in the Calabi-Yau $Y$. \Eg, $V=v_3=u_1=0$. Computing from these the intersection matrix $\mathcal{M}_4$ we find 
\be
\Gamma = \mathbb{Z}_8 \oplus \mathbb{Z}_2 \,.
\ee

%
\providecommand{\href}[2]{#2}\begingroup\raggedright\endgroup

\end{document}